\documentclass[a4paper,usenatbib]{mnras} 

\usepackage{newtxtext,newtxmath}

\usepackage[T1]{fontenc}
\usepackage{ae,aecompl}

\usepackage{amsmath} 
\usepackage{amssymb} 
\usepackage{color} 
\usepackage{graphicx} 
\usepackage[utf8]{inputenc}
\usepackage{qtree}
\usepackage[normalem]{ulem}

\newlength{\figsize}
\newlength{\figdblsize}
\newcommand{\lgal}{{\sc L-Galaxies}}

\newcommand{\Mbulge}{\mbox{$M_\mathrm{b}$}}
\newcommand{\Mdisc}{\mbox{$M_\mathrm{d}$}}
\newcommand{\Mstar}{\mbox{$M_\mathrm{d,\star}$}}
\newcommand{\Mgas}{\mbox{$M_\mathrm{d,gas}$}}
\newcommand{\Mtot}{\mbox{$M_\mathrm{d,total}$}}
\newcommand{\Msun}{\hbox{$\rm\, M_{\odot}$}}

\newcommand{\Rbulge}{\mbox{$R_\mathrm{b}$}}
\newcommand{\Rdisc}{\mbox{$R_\mathrm{d}$}}
\newcommand{\Rstar}{\mbox{$R_\mathrm{d,\star}$}}
\newcommand{\Jstar}{\mbox{$J_\mathrm{d,\star}$}}
\newcommand{\Rgas}{\mbox{$R_\mathrm{d,gas}$}}
\newcommand{\Jgas}{\mbox{$J_\mathrm{d,gas}$}}
\newcommand{\SAM}{SA model}

\newcommand{\Vvir}{\mbox{$V_\mathrm{c}$}}

\newcommand{\estar}{\mbox{$\epsilon_\mathrm{\star}$}}
\newcommand{\egas}{\mbox{$\epsilon_\mathrm{gas}$}}
\newcommand{\etot}{\mbox{$\epsilon_\mathrm{total}$}}

\newcommand{\Eq}[1]{Equation~\ref{eq:#1}}
\newcommand{\Fig}[1]{Fig.~\ref{fig:#1}}
\newcommand{\Sec}[1]{Section~\ref{sec:#1}}

\definecolor{orange}{RGB}{255, 102, 0}

\title[Morphological evolution and galactic sizes]{Morphological evolution and galactic sizes in the \lgal\ \SAM}

\author[Irodotou et al.]
{{Dimitrios Irodotou$^1$}\thanks{E-mail: Dimitrios.Irodotou@sussex.ac.uk},
 {Peter A.~Thomas$^1$}, 
 {Bruno M.~Henriques$^2$},
 {Mark T.~Sargent$^1$},
 \newauthor
 {Jessica M.~Hislop$^3$}\\
 {}$^1$Astronomy Centre, University of Sussex, Falmer, Brighton BN1 9QH,
 UK\\
 {}$^2$Institute for Astronomy, Department of Physics, ETH Zurich, 8093 Zurich, Switzerland\\
 {}$^3$Max-Planck-Institut für Astrophysik, Karl-Schwarzschild-Str. 1, D-85741 Garching, Germany\\}
\begin{document}

\pagerange{\pageref{firstpage}--\pageref{lastpage}} \pubyear{2018}

\maketitle

\label{firstpage}
\begin{abstract}
In this work we update the \lgal\ semi-analytic model (SAM) to better follow the physical processes responsible for the growth of bulges via disc instabilities (leading to pseudo-bulges) and mergers (leading to classical bulges). We address the former by considering the contribution of both stellar and gaseous discs in the stability of the galaxy, and we update the latter by including dissipation of energy in gas-rich mergers. Furthermore, we introduce angular momentum losses during cooling and find that an accurate match to the observed correlation between stellar disc scale length and mass at $z \sim 0.0$ requires that the gas loses 20$\%$ of its initial specific angular momentum to the corresponding dark matter halo during the formation of the cold gas disc. We reproduce the observed trends between the stellar mass and specific angular momentum for both disc- and bulge-dominated galaxies, with the former rotating faster than the latter of the same mass. We conclude that a two-component instability recipe provides a morphologically diverse galaxy sample which matches the observed fractional breakdown of galaxies into different morphological types. This recipe also enables us to obtain an excellent fit to the morphology-mass relation and stellar mass function of different galactic types. Finally, we find that energy dissipation during mergers reduces the merger remnant sizes and allows us to match the observed mass-size relation for bulge-dominated systems.
\end{abstract}


\section{Introduction} \label{sec:Intro}
In the widely accepted Lambda Cold Dark Matter ($\Lambda$CDM) scenario the baryonic matter collapses into the centres of dark matter haloes, where it tends to form rotationally supported discs \citep{BFP84,P84}. In this framework, dark matter haloes acquire their angular momentum through tidal interactions \citep{P69,W84,BE87}, and the associated gas discs were originally assumed \citep{FE80} to obtain the same specific angular momentum. Eventually, the collapsed gas will form stars and subsequently galaxies \citep{WR78}. While these dark matter structures evolve over time, they grow in mass and size through accretion and/or repeated mergers \citep{LC93}. Since galaxy formation occurs within haloes, these phenomena also affect the properties of the associated galaxies.

This galaxy formation paradigm has been successfully captured by semi-analytic models (SAMs), which utilise N-body simulations of dark matter to obtain information about haloes's properties, substructures and merger history, while analytic recipes infer the properties of galaxies hosted within those structures.

\subsection{Mergers and disc instabilities} \label{sec:Intro:DI}
In the hierarchical picture of structure formation, galaxy mergers have the ability to redesign the morphology of the progenitors
\citep{TT72,BH96,HBC10}. In particular, major mergers \citep{KWG93,BCF96} or multiple minor mergers \citep{BJC07} are considered to be the natural culprits for converting the stellar orbits from circular to random, hence forming spheroid-like components (i.e., classical bulges) and dispersion-supported galaxies.
 
In addition to mergers, internal secular processes \citep[see][for a review]{S13}, such as the formation and evolution of bars, \citep{CDF90,RSJ91,MSH06} are also known to be drivers of galactic evolution. Bars induce torques into discs that cause considerable outward angular momentum transfer and redistribution/migration of stellar material \citep[e.g.,][]{H71,DMC06,MF10}. Furthermore, they funnel gas to the centre of the galaxy \citep{CS81,PN90,ES04}, thus enhancing central star formation \citep{HML86,FB95,J06,HSS15}. 

In specific cases \citep{HCY09,SBW09,GBB09,GMC13} mergers may as well constitute a mechanism able to trigger gravitational instabilities \citep[e.g.,][]{T64,OP73,ELN82} and create inner disc structures \citep{ABP01,EBA06,EGB11} or starbust activity \citep{MH94}. These secular processes will culminate in the formation of a component called pseudo-bulge \citep[see][and references therein]{KK04}.

Since both stellar and gaseous discs contribute to the stability of the galaxy \citep[e.g.,][]{JS84,JS84b,BR88,WS94,E95,J96,R01,RW11}, various theoretical and/or observational studies analysed local instabilities \citep{T64} of composite discs. This dictates that modelling attempts should also follow the same path.

\subsection{Bulges: classical and pseudo} \label{sec:Intro:Bulges}
Numerous authors have investigated whether the aforementioned bulge formation scenarios lead to different bulge types with distinct intrinsic properties \citep[e.g.,][]{FD16}.
Although some more recent studies \citep[e.g.,][]{A05,FAB15,SS18} divide bulges into more categories, most authors distinguish two major types: pseudo and classical. In fact, \cite{AS94,APB95,C99} studied early- and intermediate-type spiral galaxies and concluded that bulges fall into two categories: those that can be described by an exponential profile and those by an $r^{1/4}$ profile. More recently, \cite{FD08} analysed the Sérsic index of pseudo- and classical bulges and found that 90$\%$ of the former have $n_\mathrm{b}$ < 2 and all of the latter $n_\mathrm{b}$ > 2. Moreover, \cite{CJB96} used a bulge-to-disc decomposition to calculate the ratio between bulge and disc scale lengths and concluded that their observations (correlated B/D scale lengths) strongly support a secular evolution model in which bulges with exponential surface brightness profiles emerge via disc instabilities. Additionally, \cite{F06,FDF09} compared the profile of star formation in pseudo- and classical bulges and concluded that their stars are formed via different mechanisms.

It becomes apparent that there is a lot of evidence suggesting that this dichotomy can reveal a diversity in bulge properties. This motivated us to investigate these two distinct bulge types (see \Sec{Model:Bulges}) and study their scaling relations (see \Sec{Results:Size:DSL} and \Sec{Results:MM}).

\subsection{Previous modelling work} \label{sec:Intro:PrevModel}
\subsubsection{The angular momentum of baryons} \label{sec:Intro:PrevModel:AM}
The majority of analytic and semi-analytic models of galaxy formation \citep[e.g.,][]{SP99,CLB00,MFT07,GWB11,CST16,LTR18} follow \cite{FE80} and compute disc sizes based on the assumptions that a) the cold gas disc inherits the specific angular momentum of the dark matter halo in which it forms, and b) the gas conserves its angular momentum while cooling \citep[e.g.,][]{CAF94,DSS97,MMW98}. Even though modelling explicitly these processes in semi-analytic models remains a challenging task, in this work we attempt to include this phenomenon (i.e., angular momentum losses during cooling) in the \lgal\ model (see \Sec{Model:Discs} and \Sec{A:AM}) and investigate its direct impact on galactic properties (see \Sec{Results}).

\subsubsection{Disc instabilities}
In the \lgal\ \footnote{http://galformod.mpa-garching.mpg.de/public/LGalaxies/} SAM stellar disc instabilities are identified by the \cite{ELN82} criterion, which describes baryonic discs whose self-gravity dominates; thus are unstable to global perturbations (i.e., growth of bar-like modes). Their work was limited to idealised systems, which are significantly different than those found in nature or in more detailed simulations \citep[see e.g.,][for a discussion on this topic]{A08,S13,FBB18}. However, the work of \cite{ELN82} provides a simple instability criterion which is suitable to be used by SAMs where discs are formed under similar assumptions (see \Sec{Model:Discs}). 

Determining which systems develop instabilities \citep[e.g.,][]{DH08,CEB10}, which galactic components are involved in them \citep[e.g.,][]{CST16,GCV17} and how stability is restored \citep{BBM06,MFT07,DH08,GCP15} is still an open question. It is known that gaseous discs have an influential role in galactic dynamics which becomes apparent when one considers their contribution to disc stability. Nevertheless, past modelling works have relied on a one-component stability criterion \citep[e.g.,][]{CLB00,DSW06,GWB11,HWT15}.

\subsection{The \lgal\ model} \label{sec:Intro:LGal}
The most recent version of the \lgal\ model \citep[][hereafter HWT15]{HWT15} invokes a simple argument to address disc instabilities and identify the stellar mass that has to be put into the bulge. It only takes into account the stability of the stellar disc and, as a consequence, underestimates disc instabilities and fails to reproduce the observed morphological fraction of galaxies (see \Sec{Results:GalMorph}). In this work we extend the \cite{ELN82} criterion to include cases where both stars and gas are present and investigate the contribution of gas in galactic stability.

Furthermore, the half-mass radius of classical bulges is calculated via energy conservation and the virial theorem, as described in \cite{GWB11}. This approach overestimates the size of bulges, which can be remedied by considering dissipation during gas-rich mergers (see \Sec{Model:Classical} and \Sec{Results:Size:ETGs}).

\subsection{Outline of the paper} \label{sec:Intro:Outline}
This paper is organised as follows. In \Sec{Model} we briefly describe a few vital processes regarding the \lgal\ model's approach to simulate the formation and evolution of galaxies. In addition, we present the new merger remnant size and disc instability recipes we included in our model. \Sec{Results} contains the results and \Sec{Conc} our conclusions.

\section{The model} \label{sec:Model}
The \lgal\ semi-analytic model has been well-described in the literature and we refer the reader to HWT15 for more details. Here, we briefly explain some key processes that are relevant to the purpose of this study and introduce: angular momentum losses during cooling and the updated disc instability and merger remnant size recipes.
 
We use merger trees derived from the Millennium \citep{SWJ05} simulation, which
has been shown to produce accurate properties for galaxies with stellar masses $> 10^{9} \Msun$ \citep[see][for more details]{GWB11}. The cosmological parameters used throughout this work are adopted from \cite{P14}: $\sigma_{8}=0.829$, $H_{0}=67.3\; \mathrm{km/s/Mpc}$, $\Omega_{\Lambda}=0.685$, $\Omega_{\mathrm{m}}=0.315$, $\Omega_{\mathrm{b}}=0.0487\; (f_{\mathrm{b}}=0.155)$ and $n=0.96.$
 
\subsection{Formation and properties of gaseous and stellar discs} \label{sec:Model:Discs}
As haloes form and grow, they are assigned a cosmic abundance of diffuse primordial gas, which is assumed to be shock-heated to the virial temperature \citep{RO77,S77,WR78}. That gas will either cool immediately and be added to the cold gas disc of the central galaxy, or form a quasi-static hot atmosphere and accrete onto the disc at a slower pace (see Section S1.4 of HWT15).

Hitherto the \lgal\ model followed the two core assumptions of \cite{FE80}, namely: a) baryons and dark matter acquire identical specific angular momentum distributions and b) the former conserve their angular momentum while cooling. In this work, we assume that the initial specific angular momentum of the cold gas is a fraction $f = 0.8$ of the specific angular momentum of the halo within which it is embedded (see also \Sec{A:AM}). As discussed by \cite{DB12} angular momentum losses during cooling results from the fact that the mass in $\Lambda$CDM haloes is more centrally concentrated than the angular momentum and the fact that cooling is an inside-out process (i.e., inner regions cool before the outer ones). Previous studies \citep[][]{DB12,JDK18} reported that the average value of the angular momentum retention factor is $\sim$ 0.6. We choose a slightly higher value since these simulations include additional mechanisms, such as dynamical friction, which can further reduce the specific angular moment of baryons. In addition, our choice of $f = 0.8$ provides an excellent fit \footnote{The exact value of $f$ required to reproduce the observed morphologies is potentially affected by the last term in \Eq{DeltaJgas} since its simplistic assumption for the orientation of satellite's specific angular momentum can lead to the over-prediction of the specific angular momentum of the gaseous disc.} to the galactic morphologies as we discuss in \Sec{Results:GalMorph}. Finally, we note that the angular momentum loss should be transmitted to the dark matter but this effect will be relatively small and we choose to neglect it.

As discussed in \cite{GWB11}, there are three mechanisms capable of altering the angular momentum vector of the gaseous disc, namely the addition of gas by cooling, the removal of gas through star formation and the accretion from minor mergers. These can be expressed mathematically by the following formula:
\begin{align} \label{eq:DeltaJgas}
\delta \vec{J}_\mathrm{d,gas} &= \delta \vec{J}_\mathrm{gas,cooling} + \delta
\vec{J}_\mathrm{gas,SF} + \delta \vec{J}_\mathrm{gas,acc.} \nonumber \\ &=
f \frac{\vec{J}_\mathrm{DM}}{M_\mathrm{DM}} \dot{M}_\mathrm{cool} \, \delta t
\nonumber \\ &-\, \frac{\vec{J}_\mathrm{d,gas}}{\Mgas} ((1 - R_\mathrm{ret})
\dot{M}_{\star} \, \delta t + \delta M_\mathrm{reheat}) \nonumber \\ &+
\frac{\vec{J}_\mathrm{DM}}{M_\mathrm{DM}} M_\mathrm{gas,sat}\; ,
\end{align}
where the factor $f=0.8$ accounts for angular momentum losses during cooling, $\dot{M}_\mathrm{cool}$ is the cooling rate (see Equation S6 and S7 of HWT15), $\delta t$ is the time interval, $(1 - R_\mathrm{ret})
\dot{M}_\mathrm{\star}$ is the formation rate of long lived stars (see Equation S14 of HWT15), $\delta M_\mathrm{reheat}$ is the cold gas reheated into the hot atmosphere as a result of star formation activity (see Equation S18 of HWT15) and $M_\mathrm{gas,sat}$ is the cold gas mass of the merging satellites.

Following \cite{MMW98}, we assume that the gaseous and stellar discs are infinitesimally thin, in centrifugal equilibrium and have exponential surface density profiles, hence their scale-lengths can be written as:
\begin{align} \label{eq:Rg}
\Rgas &= \frac{\Jgas}{2 V_{\rm{max}} \, \Mgas}\; , \\
\Rstar & = \frac{\Jstar}{2 V_{\rm{max}} \, \Mstar}\; , \label{eq:Rd}
\end{align}
where $\Jgas$, $\Mgas$ and $\Jstar$, $\Mstar$ are the angular momentum and mass of the gaseous and stellar disc, respectively and $V_{\rm{max}}$ is the maximum circular velocity of their host halo which is used as a proxy for the rotation velocity of both discs.

\subsection{Disc instabilities} \label{sec:Model:DI}
Disc instabilities describe systems where the attraction due to self-gravity
overcomes the centrifugal force due to rotation. In our updated instability
recipe we treat galactic discs as two-component systems where the contribution
of each disc (i.e.~stellar and gaseous) to the stability of the whole galaxy is
mass-weighted. We extend the simple criterion of \cite{ELN82} and combine both
discs in an approach similar to the one used by authors who combined the
\cite{T64} local instability criteria \citep[e.g.~][]{WS94,RW11}. Our new
galactic instability criterion can be written as $\etot<1$, where:
\begin{equation} \label{eq:DI}
\Mtot\etot \equiv \Mstar\estar + \Mgas\egas \;.
\end{equation}
Here $\Mtot$, $\Mstar$ and $\Mgas$ are, respectively, the total disc mass, the mass in stars and the mass of gas in the disc, and:
\begin{equation} \label{eq:DIcomp}
\epsilon_{i} = c_i\left(G\Mdisc_{,i}\over\Vvir^2\Rdisc_{,i}\right)^{1\over2}\;,
\end{equation}
where the index $i \mathrm{= \star\; or \; gas}$, $\Vvir$ is the rotational
velocity which for both discs is approximated by the circular velocity of their
host halo, and $\Mdisc_{,i}$ and $\Rdisc_{,i}$ are the mass and scale length of the $i$ component. $c_\star=1.1$ and $c_\mathrm{gas}=0.9$
are constants that reflect the stability criteria for isolated stellar
\citep{ELN82} and gaseous \citep{CST95} discs, respectively.

If the galaxy is unstable then we adopt the following 2-step procedure:
\begin{itemize}
\item If $\egas>1$, then we transfer mass from the gas disc to the stellar disc,
  thus lowering $\egas$, until either the combined system becomes stable,
  $\etot=1$, or $\egas=1$. When making this transfer, we assume that a small
  fraction of the gas makes its way onto the central black hole in accordance with
  Equation~S23 of HWT15 (and setting the factor
  $M_\mathrm{sat}/M_\mathrm{cen}=1$ in that equation).
\item If the system remains unstable, then we transfer stars from the stellar
  disc to the bulge until $\etot=1$.
\end{itemize}

We expect that disc instability will occur mostly in the inner regions of the galaxy in which stars have low angular momentum. For that reason, when transferring stars from the disc to the bulge, we assume that they carry no angular momentum with them\footnote{A later version of the \lgal\ model, in development, will have spatially-resolved discs and be able to investigate this in more detail.}. In our model, that then results in an increased specific angular momentum of the stars left behind and (see \Eq{Rd}) a proportional increase in the disc scale length.

\subsection{Formation and properties of classical and pseudo-bulges} \label{sec:Model:Bulges}
In the L-Galaxies model bulges form through three distinct mechanisms: major mergers, minor mergers and disc instabilities. Major and minor mergers are assumed to produce classical bulges, while disc instabilities lead to the formation of bar-related pseudo-bulges.

\subsubsection{Classical bulges} \label{sec:Model:Classical}
Whenever two or more dark matter haloes merge, so do their associated galaxies but on a longer timescale determined by 2-body relaxation. In our model we characterise as central galaxies those that dwell in the potential minimum of the most massive subhaloes (hereafter the main halo) and as satellite galaxies those that reside inside the non-dominant subhaloes that are bound to the main halo.

We distinguish a major from a minor merger based on the ratio of the total baryonic mass (stars+gas), $M_1$ and $M_2$, of the satellite and central galaxy, respectively. In major mergers ($M_1/M_2 > 0.1$, see HWT15 for more details) the discs of the progenitors are dismantled and both the pre-existing stars and those formed during the merger become part of the resulting bulge-dominated galaxy. In minor mergers, the bulge of the descendant accretes all the stars of the less massive progenitor, while stars formed during this process remain in the remnant's disc. The mass of those stars is calculated by using the ``collisional starburst”
formulation of \cite{SPF00}:
\begin{equation} \label{eq:Mstarburst}
M_\mathrm{\star,burst} = \alpha_\mathrm{SF,burst} \left(
\frac{M_1}{M_2}\right)^{\beta_\mathrm{SF,burst}} \Mgas\; ,
\end{equation}
where $\alpha_\mathrm{SF,burst}$ and $\beta_\mathrm{SF,burst}$ are free parameters and $\Mgas$ is the total gas disc mass of both galaxies combined.

Galaxy mergers are considered to play a fundamental role in the production of elliptical galaxies, hence having a model able to evaluate the size of the remnant and reproduce its scaling relations across cosmic time is crucial. The HWT15 version of the model calculated the half-mass radius of the remnant using energy conservation arguments, where the final binding energy was equated to the self-binding energies of the two progenitors plus their orbital energy (see Equation S34 of HWT15). Several authors have argued that this simple picture leads to unrealistic sizes, especially at the low-mass end
\citep{CDC07,HHC09,SMB10,HBH10,CPP11,SMB13,PSP14}. This problem mainly arises from the fact that the above approximation does not take into account gas dissipation during mergers, where gas clouds collide and radiate away their kinetic energies. In cases where the gas makes up a significant fraction of the total mass of the progenitors this phenomenon would result in smaller and denser remnants. We follow \cite{CDC07,CPP11} and \cite{TMC16} and include a term to account for radiative losses. In this picture the energy conservation formula is given by:
\begin{equation} \label{eq:energycon}
E_\mathrm{final} = E_\mathrm{initial} + E_\mathrm{orbital} +
E_\mathrm{radiative}\; ,
\end{equation}
where for major mergers each energy term can be explicitly written as:
\begin{align}
E_\mathrm{final} &= -G \left[ \frac{(M_{\star,1} + M_{\star,2} +M_\mathrm{\star,burst} )^{2}}{R_\mathrm{final}} \right]\;
, \label{eq:finalenergyM} \\ E_\mathrm{initial} &= -G \left(
\frac{M_{1}^{2}}{R_{1}} + \frac{M_{2}^{2}}{R_{2}} \right) \;
, \label{eq:initialenergyM} \\ E_\mathrm{orbital} &= -G\left( \frac{M_{1} \, M_{2}}{R_{1} + R_{2}}\right) \; , \label{eq:orbitalenergy}
\\ E_\mathrm{radiative} &= -C_\mathrm{rad} \, E_\mathrm{initial} \left(
\frac{M_\mathrm{gas,1} + M_\mathrm{gas,2}}{M_{1} + M_{2}} \right)\;
, \label{eq:radenergyM}
\end{align}
where $M_\mathrm{\star,i}$, $M_\mathrm{i}$, $M_\mathrm{gas,i}$ and $R_\mathrm{i}$ are the total stellar mass (disc+bulge), total baryonic mass (stars+gas), gas mass and stellar half-mass radius of the \textit{i} progenitor, $R_\mathrm{final}$ is the stellar half-mass radius of the remnant, $C_\mathrm{rad}$ is a parameter which defines the efficiency of the radiative process (see discussion below) and $M_\mathrm{\star,burst}$ is the mass of the new stars formed during the merger which is given by \Eq{Mstarburst}.

For minor mergers we follow \cite{GWB11} and assume that the total stellar mass of the satellite galaxy is merged with the bulge of the central galaxy, therefore:
\begin{align}
E_\mathrm{final} &= -G \left[ \frac{(M_\mathrm{b,1} + M_{\star,2})^{2}}{R_\mathrm{final}} \right]\;
, \label{eq:finalenergy} \\ E_\mathrm{initial} &= -G \left(
\frac{M_\mathrm{b,1}^{2}}{R_\mathrm{b,1}} + \frac{M_{\star,2}^{2}}{R_{2}} \right) \;
, \label{eq:initialenergy} \\ E_\mathrm{orbital} &= -G\left( \frac{M_\mathrm{b,1} \, M_{\star,2}}{R_\mathrm{b,1} + R_{2}}\right) \; , \label{eq:orbitalenergy}
\\ E_\mathrm{radiative} &= -C_\mathrm{rad} \, E_\mathrm{initial} \left(
\frac{M_\mathrm{gas,1} + M_\mathrm{gas,2}}{M_{1} + M_{2}} \right)\;
, \label{eq:radenergy}
\end{align}
where $M_\mathrm{b,1}$ and $R_\mathrm{b,1}$ are the stellar mass and half-mass radius of the bulge of the more massive progenitor and $M_{\star,2}$ and $R_{2}$ are the total stellar mass and half-mass radius of the minor progenitor. \Eq{energycon} indicates that galaxies with higher gas fractions will experience more dissipation during mergers, and since lower mass galaxies have low-mass progenitors which have higher cold gas fractions at all redshifts, early-wet mergers will produce more compact remnants than late-dry mergers \citep[e.g.][]{RBC06,RCH06,DC06}.

\cite{CDC07} calibrated their model using the N-Body/SPH code GADGET \citep{SYW01} to simulate mergers of isolated, low-redshift, gas-rich, identical disc galaxies, finding $C_\mathrm{rad}\simeq$1. However, a higher value of $C_\mathrm{rad} = 2.75$ was found for disc-dominated galaxies that have recently experienced a major merger \citep{CPP11}. In addition, \cite{PSP14} used 68 hydrodynamical simulations of major and minor binary mergers \citep[see][]{JNB09} of galaxies with either mixed or spheroid-spheroid morphologies. They found that the morphology, the mass ratio and the gas content could cause the $C_\mathrm{rad}$ parameter to vary significantly, from 0.0 (dissipationless) for minor or major mergers where one or both of the progenitors are bulge-dominated; to 2.5 for major mergers between disc-dominated galaxies. In the current work we adopt the value $C_\mathrm{rad} = 1.0$ in concordance with previous modellers.
 
\subsubsection{Pseudo-bulges} \label{sec:Model:Pseudo}
Galaxy-galaxy interactions have a pivotal role in regulating galactic evolution, however internal processes, such as disc instabilities, are of similar importance since they are responsible for the emergence of pseudo-bulges.

In order to determine the half-mass radius of the resulting pseudo-bulge we distinguish between two cases; the first is when the disc already possesses a bulge and then becomes unstable. We follow \cite{GWB11} and assume that the unstable mass merges into the existing bulge, thus the final bulge's half-mass radius is given by:
\begin{equation} \label{eq:DIBulgesize}
C \frac{G M_\mathrm{final}^{2}}{R_\mathrm{final}} = C
\frac{G M_\mathrm{old}^{2}}{R_\mathrm{old}} + C
\frac{G M_\mathrm{\star,unstable}^{2}}{\Rbulge} + \alpha_\mathrm{inter} 
\frac{G M_\mathrm{old} \, M_\mathrm{\star,unstable}}{R_\mathrm{old} + \Rbulge}\; ,
\end{equation}
where $M_\mathrm{final}$ and $R_\mathrm{final}$ are the mass and half-mass radius of the final bulge, $M_\mathrm{old}$ and $R_\mathrm{old}$ are the mass and half-mass radius of the existing bulge, $M_\mathrm{\star,unstable}$ is the mass occurred from \Eq{DI}, C is a structural parameter relating the binding energy of a galaxy to its mass and radius and $\alpha_\mathrm{inter}$ is a parameter quantifying the effective interaction energy deposited in the stellar components. \cite{GWB11,HWT15} used $\alpha_\mathrm{inter} / C = 4$ as it led to bulge sizes in adequate agreement with SDSS galaxies. However, we set $\alpha_\mathrm{inter} / C = 1.5$ in order to be consistent with the results of \cite{BMQ05} who found that $1.3 < \alpha_\mathrm{inter} / C < 1.7$. The half-mass radius of the unstable material $\Rbulge$ is taken from:
\begin{align} \label{eq:UnstableStars}
M_\mathrm{\star,unstable} &= 2 \pi \, \Sigma_{\star,0} \, \Rstar \nonumber
\\ & \, \left[ \Rstar\, -\, (\Rbulge + \Rstar) \, \mathrm{exp}(-\Rbulge/\Rstar) \right] \; ,
\end{align}
where $\Sigma_{\star, 0}$ and $\Rstar$ are the central surface density and the exponential scale length of the disc. If the galaxy had no bulge prior to the instability incident we assume that the half-mass radius of the newly formed bulge is equal to $\Rbulge$.

Finally, in the \lgal\ SAM the specific angular momentum of bulges is only
altered during mergers, since we assume that during instabilities the unstable
stellar mass transfers negligible angular momentum from the disc to the bulge
\citep[as in][]{GWB11,HWT15}. We assume that the accreted specific angular
momentum matches that of the halo within which the satellite galaxy is orbiting,
such that the specific angular momentum of the merger remnant can be written as:
\begin{equation} \label{eq:DIBulgespin}
j_{\mathrm{b}} = \frac{j_{\mathrm{b,old}} \, M_{\mathrm{old}} + j_{\mathrm{halo}} \, M_{\mathrm{\star,sat}}}{M_{\mathrm{b}}}\; ,
\end{equation}
where $j_{\mathrm{b,old}}$ and $M_\mathrm{old}$ are the specific angular
momentum and mass of the existing bulge, $j_{\mathrm{halo}}$ is the specific
angular momentum of the halo, $M_\mathrm{\star,sat}$ is the stellar mass of the
satellite and $M_\mathrm{b}$ is the remnant's new mass.

\section{Results} \label{sec:Results}
\subsection{Stellar mass assembly channels} \label{sec:Results:Decomposition}
\begin{figure*}
\centering
\includegraphics[width=\figdblsize]{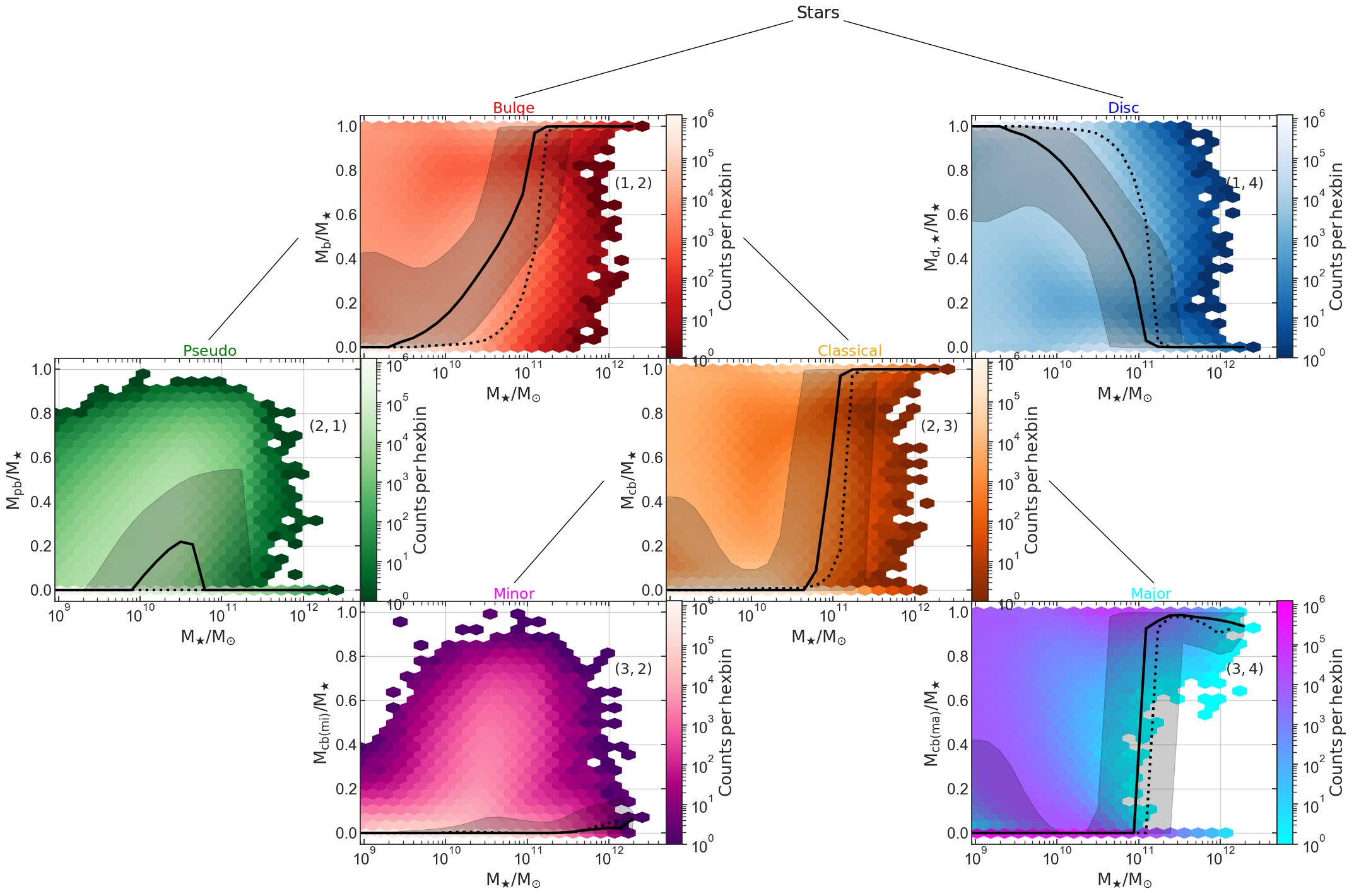}
\caption{Decomposition of the total stellar mass of each galaxy at redshift z $\sim$
  0.0. Each panel contains a hexagonal binning plot of the component-to-total
  stellar mass ratio as a function of the total stellar mass along with our (black solid) and HWT15 (black dotted) median lines and our $\mathrm{16^{th}-84^{th}}$ percentile range
  (black shaded regions). The black straight lines that connect the panels
  represent the divisions described in the tree chart in
  \Sec{Results:Decomposition} and the (row, column) positioning of each
  component in the figure corresponds to: (1,2) -- bulge; (1,4) -- disc; (2,1)
  -- pseudo-bulge; (2,3) -- classical-bulge; (3,2) -- through minor mergers;
  (3,4) -- through minor mergers.}
\label{fig:SM_Decomp}
\end{figure*}

One feature of our model is its ability to follow the formation and evolution of
classical and pseudo-bulges by separately tracking each channel that contributes to their mass budget. This allows us to gain insight into the behaviour of each component and answer questions such as: how often do disc galaxies host classical as opposed to pseudo-bulges; how is the mass of disc- and bulge-dominated galaxies distributed; and how structurally different are galaxies that host classical or pseudo-bulges.

We follow the stellar mass transferred between galaxies in minor and major
mergers, and the stellar mass transferred between galactic components during
disc instability events. We split the total stellar mass into 6 categories, some
of which are subsets of others: \\ \qtreecenterfalse \Tree [.Stars-M_{\star} [
    \textcolor{green}{Pseudo}-M_{\mathrm{pb}} [
      \textcolor{magenta}{Minor}-M_{\mathrm{cb(mi)}} !\qsetw{3cm}
      \textcolor{cyan}{Major}-M_{\mathrm{cb(ma)}}
    ].\textcolor{orange}{Classical}-M_{\mathrm{cb}}
  ].\textcolor{red}{Bulge}-M_{\mathrm{b}} !\qsetw{3cm}
  \textcolor{blue}{Disc}-M_{\mathrm{d,\star}} ]
\newline

In the \lgal\ model stars can be found in the two main galactic components,
namely the stellar disc (d,$\star$) and the bulge (b). Tracking the two widely accepted bulge formation paths allows us to further divide the bulge mass into the mass created via disc instabilities and the one accreted through mergers, hence leading to the formation of pseudo-bulges (pb) and classical bulges (cb), respectively. Finally, the population of classical bulges can be dichotomised into those produced through major mergers (cb(ma)) and those through minor mergers (cb(mi)). This decomposition is shown in \Fig{SM_Decomp} which contains the ratio between the stellar mass of each of the above 6 components and the total stellar mass of each galaxy. 

As can be seen in \Fig{SM_Decomp}, the \lgal\ model produces pure disc- (1,4) with masses up to $M_{\star} \sim 3 \, 10^{11} \Msun$ and pure bulge-dominated galaxies (1,2) of all masses. The corresponding median lines suggest that the most massive galaxies are bulge-dominated \citep[e.g.,][]{BGB04,WE12,NA10}, while the majority of normal galaxies are
disc-dominated \citep[e.g.,][]{FNO07,BNB09,NA10}; a behaviour which is consistent with observational studies. 

The large scatter in (2,1) suggests that in a few galaxies pseudo-bulges dominate the total stellar mass budget, hence leading to the development of lenticular galaxies \citep{KK04,KC04,WJK09,VBK13}. Interestingly, there are a some extreme cases where the pseudo-bulge-to-total stellar mass ratio is as high as 0.9. This is in agreement with the recent work of \cite{SC18} who proposed disc instabilities as a mechanism responsible for the production of field S0 galaxies. 

From the behaviour of our data in panels (2,1) and (2,3) and
the corresponding median lines, we can say that most of the bulge mass in
galaxies with masses between $10^{10} \Msun < M_{\star} < 10^{11} \Msun$ is in pseudo- and not classical bulges. At $\sim 10^{11} \Msun$ the secular
evolution scenario cannot compete with the violent one and as a consequence
major mergers (3,4) begin to destroy the progenitors and form bulge-dominated
systems (2,3). 

Finally, we note that minor mergers (3,2) never have a
significant contribution to the total stellar mass due to the adopted merger mass ratio threshold\footnote{Set to 0.1 by HWT15 to ensure that the majority of high-mass galaxies were bulge-dominated.}. High-resolution simulations \citep[e.g.,][]{SBW08,HBC10} indicate that mergers with mass ratios 0.1 and below are very rare and have relatively little impact on the total stellar mass of the remnant galaxy. In addition, in cases where a minor merger triggers disc instabilities, we assume that the newly formed pseudo-bulge will contain both the unstable and the previously existing bulge mass. Thus, the stellar mass accreted from that minor merger is now considered to be part of the pseudo-bulge.

The most notable discrepancies between the HWT15 (dotted lines) and the presented version of the L-Galaxies SAM appear in panels (1,2), (1,4) and especially (2,1). These three panels highlight the importance of gas in disc instabilities since it enhances the formation of bar-related pseudo-bulges \citep{BC02,SKK19} and reduces the unrealistic population of high mass disc-dominated galaxies seen in HWT15.
 
\subsection{Stellar mass evolution of galactic components} \label{sec:Results:DecompositionEvo}
\begin{figure}
\centering
\includegraphics[width=\figsize]{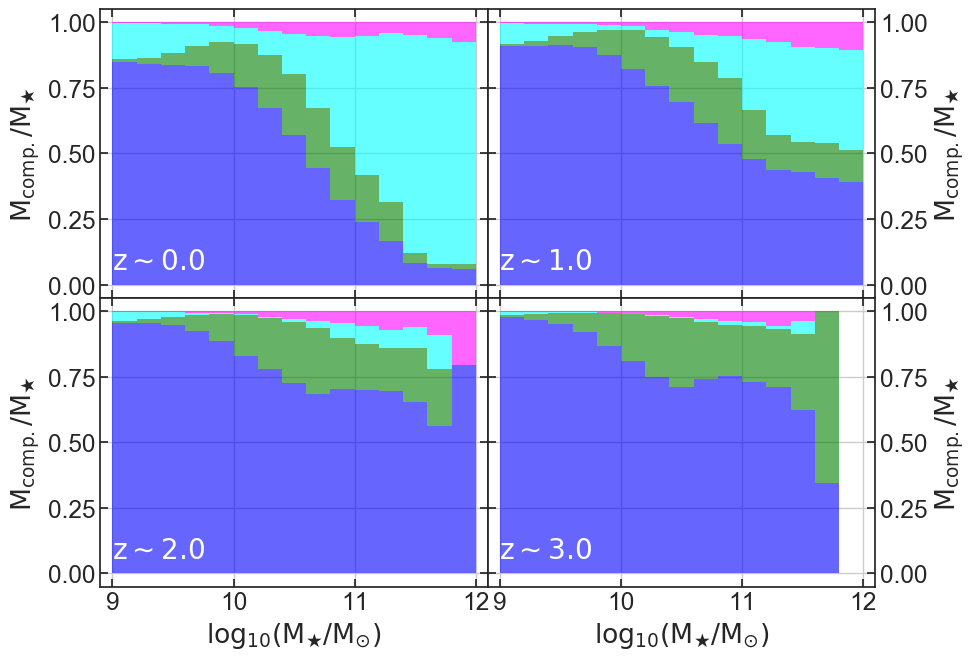}
\caption{Relative contribution of each component to the total stellar mass (i.e. the mass summed over all galaxies in a given stellar mass bin) at
  redshifts $z \sim$ 0.0, 1.0, 2.0 and 3.0. Colours are the same as in
  \Fig{SM_Decomp}: blue -- disc; green -- pseudo-bulge; cyan -- classical bulge
  through major mergers; magenta -- classical bulge through minor mergers.}
\label{fig:CompMassRatio_Vs_SM_Evo}
\end{figure}

\Fig{CompMassRatio_Vs_SM_Evo} illustrates how the contribution of each component to the total stellar mass fluctuates within each galactic mass range at different redshifts.

At $z \sim 0.0$ at the lowest masses about 80 percent of stars lie in discs with most of the rest in merger-driven bulges. At $10^{10} \Msun$ minor mergers begin to initiate the formation of classical bulges and for total stellar masses above $10^{11} \Msun$ major mergers become the dominant mechanism that affects galactic morphology. This behaviour follows from a hierarchical galaxy assembly scenario in which mergers give rise to the formation of the most massive system. Pseudo-bulges never dominate but are most important between $10^{10} \Msun$ and $10^{11.5} \Msun$, accounting for about 20 percent of the total over this range.
 
As pointed out by few authors \citep[e.g.,][]{BWM04,CBP05,ISF10} the massive end of the galaxy mass function at $z < 0.8$ is dominated by galaxies with early-type morphologies, which is consistent with our results. Furthermore, studies which focused on the evolution of the merger rate of galaxies \citep[e.g.,][]{FAL00,BPS06,LDF08} concluded that the majority of them have experienced major mergers since $z \sim 1$, and this event has severely affected their morphology \citep{D05}. In galaxies produced by the \lgal\ model we can also notice that classical bulges, both via major and minor merger, have a significant contribution to the total stellar mass at $z \leq 1$, while the disc component becomes increasingly important at higher redshifts. Finally, we see a notable evolution in the contribution of pseudo-bulges in the total stellar mass, which is linked to the high gas content of high redshift galaxies. This evolution is consistent with simulations \citep[e.g.,][]{ATM09,FKB14} and is captured by our modified disc instability recipe which takes into account the contribution of the gaseous disc in the galactic stability.

\subsection{Stellar mass functions}
\begin{figure}
\centering \includegraphics[width=\figsize]{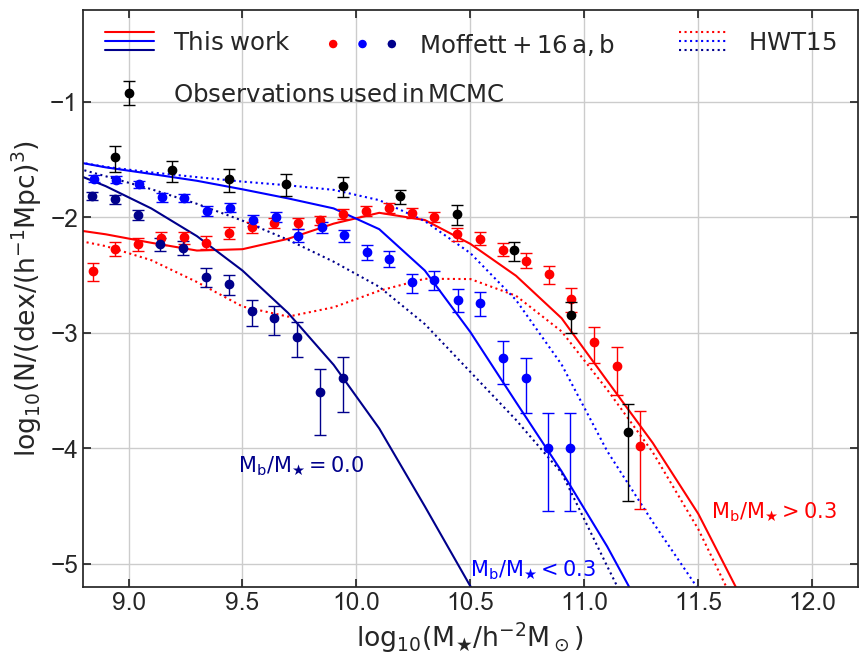}
\caption{Total stellar mass function at redshift z $\sim$ 0.0 for different galactic types. Blue, red and darkblue lines represent disc-dominated galaxies, systems with $\Mbulge / M_{\star} > 0.3$ and pure disc galaxies, respectively. Blue and red circles represent observational data from \protect\cite{MID16}, while darkblue from \protect\cite{MLD16}. Solid and dotted lines show results from this work and HWT15, respectively}
\label{fig:SMF_Morph}
\end{figure}

As discussed in \Sec{Model:DI}, stellar and gaseous discs are able to trigger instabilities that can significantly redistribute galactic material. Hence, stellar mass functions of disc and bulge stars allow us to illustrate the influence of our new approach on galaxies produced by the \lgal\ model (see also \Fig{SMF}).

In \Fig{SMF_Morph} we split our galaxies into different morphological types according to their bulge mass fractions and plot their total stellar mass functions. We compare with the \cite{MLD16} sample of single-component pure disc systems and the disc-(Sab-Scd/Sd-Irr) and spheroid-dominated (E/S0-Sa) galaxies selected by \cite{MID16}. The choice of a bulge to total mass ratio of 0.3 to distinguish disc- and spheroid-dominated systems in our model was motivated by various studies of the $B/T$ ratio of S0 galaxies \citep[e.g.][]{LSB05,LSB10,BSV16} which found that the mean value is $\sim 0.25$. In addition, \cite{WJK09} found that, in their sample, the fraction of spiral galaxies with $B/T > 0.4$ is $8\%$. Hence, our $B/T$ threshold lies between these observed values.

The updated \lgal\ model shows an impressive agreement with the behaviour denoted by the observational data for all galaxy samples. On the other hand, it is clear that the HWT15 model has a strong tendency to form more disc-dominated systems and bulgeless galaxies at higher masses. On that account, the instability recipe described in \Sec{Model:DI} prevents the formation of disc-dominated galaxies and restricts the abundance of purely disc systems by redistributing the stellar mass between galactic components. Hence, it directly affects the galactic morphology and allows us to better match the observed behaviours in \Fig{SMF_Morph}.

\subsection{Galactic morphology} \label{sec:Results:GalMorph}
\begin{figure*}
\centering \includegraphics[width=\figdblsize]{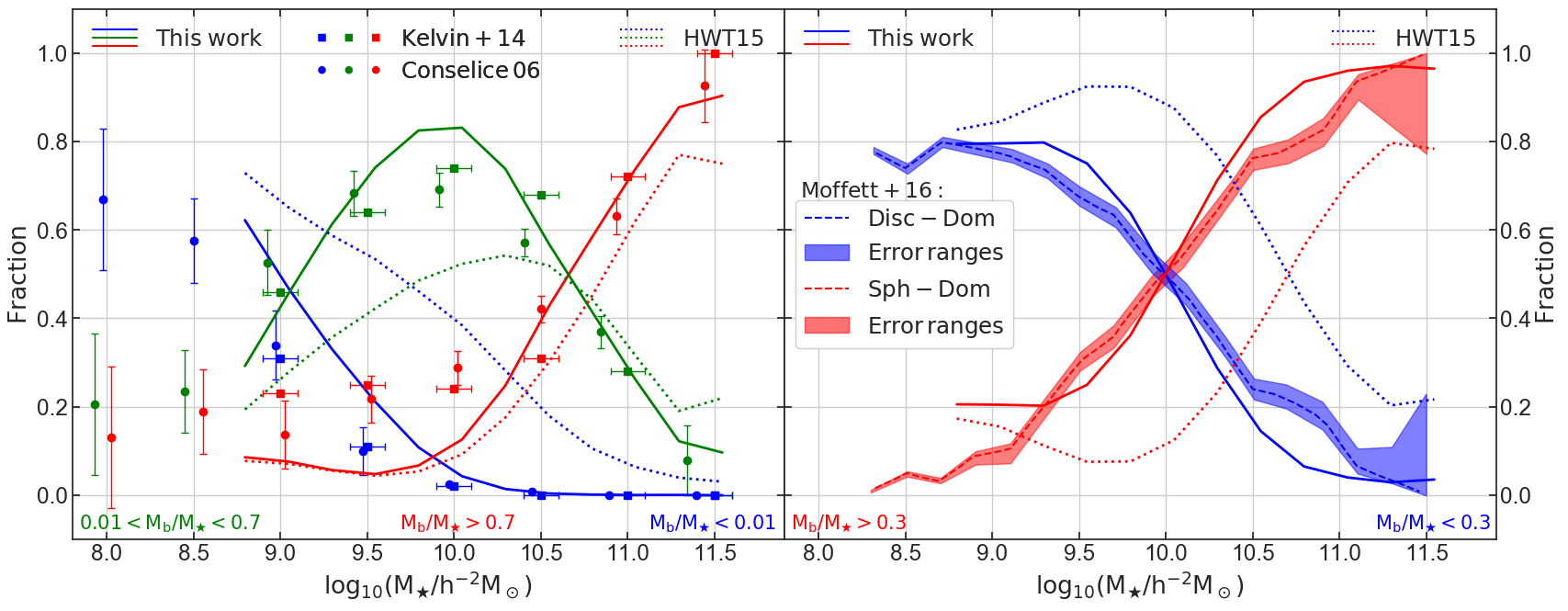}
\caption{Fraction of different morphological types as a function of total
  stellar mass at redshift z $\sim$ 0.0. \textbf{Left panel:} Red, green and blue lines
  show the fraction of bulge-dominated, normal spiral and pure disc galaxies,
  respectively. Red, green and blue filled circles are observational data from
  \protect\cite{C06} that show the elliptical, spiral and irregular galaxies,
  respectively. Red, green and blue squares are S0-Sa+Sab-Scd, LBS+E and Sd-Irr galaxies, respectively from \protect\cite{KDR14}. \textbf{Right panel:} Red lines represent systems with $\Mbulge / M_{\star} > 0.3$ while blue lines represent disc-dominated. Observational data points from \protect\cite{MID16} are represented by dashed lines along with the corresponding errors. In both panels solid and dotted lines show results from this work and HWT15, respectively}
\label{fig:Gal_Frac}
\end{figure*}

In the \lgal\ SAM galaxy mergers are dichotomised into major and minor. If the total baryonic mass (stars+gas) of the more massive
progenitor exceeds that of the less massive by at least an order of magnitude,
then this incident is characterised as minor; in any other case the merger is
treated as a major. In this work we adopt the same mass ratio threshold
($R_\mathrm{merger} \equiv M_\mathrm{sat.}/M_\mathrm{cen.} = 0.1$)\footnote{Chosen by HWT15 to ensure that the majority of galaxies above $10^{11.5} \Msun$ are bulge-dominated.} as in HWT15 in order to distinguish those two regimes. This division regulates the type of the remnant galaxy (i.e., bulge- or disc-dominated).

This categorisation is depicted in the left panel of \Fig{Gal_Frac} which represents the fraction of different galaxy types as a function of their total stellar mass. We split our galaxy sample into three categories based on their bulge-to-total stellar mass ratio. In \Fig{Gal_Frac} the
red solid line shows the fraction of bulge-dominated galaxies, akin to
ellipticals $(\Mbulge/M_{\star}>0.7)$, blue solid line shows the fraction of
normal spirals $(0.01<\Mbulge/M_{\star}<0.7)$ and green solid line represents
disc-dominated galaxies, akin to extreme late-types
$(\Mbulge/M_{\star}<0.01)$. Similar approaches for proxies for the
morphology of simulated galaxies have been used by several authors \citep[e.g.,][]{BDT07,LCP08,GWB11,GCP15}. From an observational point of view,
\cite{WJK09} measured the $B/T$ ratio of 143 bright, high mass spirals and
concluded that $\sim 66\%$, $\sim 26\%$, $\sim 8\%$ and $100\%$ of their
galaxies have $B/T \leq 0.2$, $0.2 < B/T \leq 0.4$, $0.4 < B/T \leq 0.75$ and
$B/T \leq 0.75$, respectively: throughout this work we use $\Mbulge /
M_{\star} < 0.3$ (i.e., $\Mdisc / M_{\star} > 0.7$) for disc-dominated and
$\Mbulge / M_{\star} > 0.7$ for bulge-dominated galaxies.

The observational data has been taken from \cite{C06}, who used a sample of $\sim$ 22,000 galaxies at $z < 0.05$ to plot the morphological fraction as a function of stellar mass. In addition, \cite{KDR14} analysed a sample of 2,711 local (0.025 $< z <$ 0.06) galaxies taken from the Galaxy And Mass Assembly (GAMA) survey. They visually divided their sample into 5 categories, namely LBS, E, S0-Sa, Sab-Scd and Sd-Irr, however in order to make the comparison with our data more efficient we combined the LBS with E galaxies and the S0-Sa with Sab-Scd galaxies \citep[see Table 1 of][]{KDR14}. We also include the HWT15 data (dotted lines) for comparison with the previous version of the model.

Both observational surveys indicate that the fractional contribution of galaxies with stellar masses between $10^{9} \Msun < M_{\star} < 10^{11} \Msun$ is dominated by spirals, although by $M_{\star} \sim 10^{10.5} \Msun$ spirals and ellipticals represent about 50$\%$ each. At stellar masses higher than that, almost all galaxies have turned into ellipticals. These behaviours are also fairly well represented by our galaxies over the whole stellar mass range. However, despite the new instability recipe the \lgal\ model fails to match the fraction of bulge dominated galaxies for masses below $\sim 10^{10} \Msun$. Tidally induced bars \citep[e.g.,][]{RCP15,LE16,PL19} may represent a mechanism capable of altering this behaviour by further transferring mass to the bulge; and we plan to test this effect in future work.

In the right panel we present the fraction of our disc-dominated galaxies ($\Mbulge / M_{\star} < 0.3$) and systems with $\Mbulge / M_{\star} > 0.3$ and compare with the fraction found by \cite{MID16} who selected 4,971 disc-(Sab-Scd/Sd-Irr) and 1,692 spheroid-dominated (E/S0-Sa) galaxies. We use the same selection criteria as those described in the previous section since we compare with the same survey. Our results suggest that the point indicating the transition between the numerical dominance of disc- and spheroid-dominated galaxies is in strong agreement with \cite{MID16} and shows a clear improvement over the HWT15 version of the \lgal\ \SAM.

\subsection{Mass-specific angular momentum relation}
\label{sec:Results:Spin}
\begin{figure*}
\centering \includegraphics[width=\figdblsize]{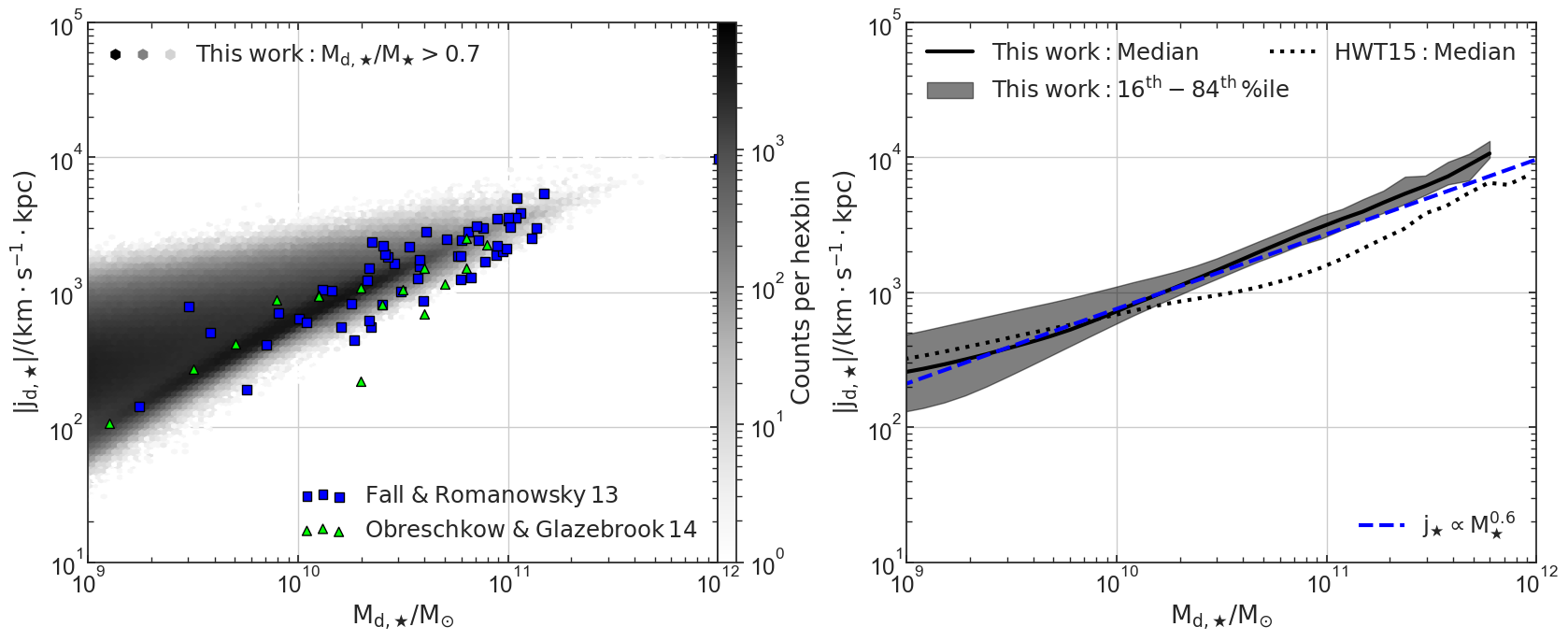}
\caption{Mass-specific angular momentum relation for disc-dominated
  galaxies at redshift z $\sim$ 0.0.
  \textbf{Left panel:} Disc stellar mass versus specific angular momentum    compared with \protect\cite{OG14} and
  \protect\cite{FR13} observations. \textbf{Right panel:} The median and
  $\mathrm{16^{th}-84^{th}}$ percentile range (black shaded region) of the aforementioned relation
  compared with the \protect\cite{F83} relation. Black solid and dotted lines show results from this
  work and HWT15, respectively.}
\label{fig:DM_Vs_DS}
\end{figure*}

\begin{figure}
\centering \includegraphics[width=\figsize]{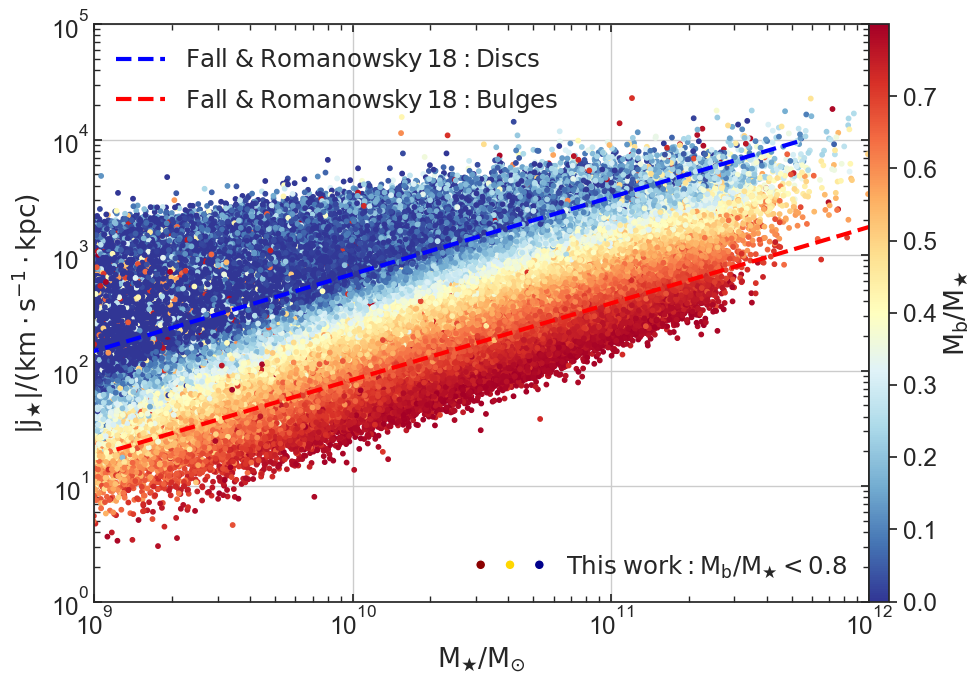}
\caption{Total stellar mass versus total specific angular momentum for $\Mbulge /
  M_{\star} < 0.8$ galaxies at redshift z $\sim$ 0.0, compared with
  \protect\cite{FR18} fit lines for discs and bulges. The colour of the symbols
  indicate different $\Mbulge / M_{\star}$ values.}
\label{fig:SM_Vs_GS}
\end{figure}

Angular momentum is one of the most fundamental galactic properties; it can dictate the galactic size and morphology, and also provides a vital constraint on theories of galaxy formation \citep[e.g.,][]{MMW98,RF12,OG14,SFG18,PFT18}. The correlation of the specific angular momentum with stellar mass, $j \propto M^{2/3}$, was introduced 35 years ago by \cite{F83}.

We show this relation in \Fig{DM_Vs_DS} for our disc-dominated galaxies ($\Mstar /M_{\star}>0.7$). We also include results from \cite{OG14} who analysed 16 nearby spiral galaxies of the The HI Nearby Galaxy Survey (THINGS) sample \citep{WBB08}, and \cite{FR13} who focused on 64 galaxies from type Sa to Sm from the \cite{K86,K87,K88} datasets, and find that our simulated galaxies follow closely the \cite{F83} relation and are in very good agreement with the observations. We also notice that for disc masses $\sim 10^{10}\Msun$ and above the differences between this work and HWT15 are mostly due to the new disc instability recipe since the formation of pseudo-bulges, which is expected to happen in this mass regime (see panel (2,1) in \Fig{SM_Decomp}), increases the specific angular momenta of stellar discs (see also \Sec{A:SD}).

In \Fig{SM_Vs_GS} we calculate the total specific angular momentum of each
galaxy as $j_{\star} = (j_{\mathrm{d,\star}}\Mstar + j_{\mathrm{b}}\Mbulge) /
(\Mstar + \Mbulge)$ and plot it as a function of the total stellar mass. The
different colours represent the $\Mbulge /M_{\star}$ ratio of the corresponding
galaxy. We compare our results with \cite{FR18} who presented their sample of 57 spirals, 14 lenticulars and 23 ellipticals. The behaviour of our data indicate that the more disc-dominated galaxies (i.e., lower $\Mbulge /M_{\star}$ values) rotate faster than the bulge-dominated, hence we find an impressive agreement with the observed trends.

\subsection{Mass-size relations}
\label{sec:Results:Size}
Galactic mass and size are amongst the most fundamental properties and modelling their relation remains an important task for SAMs \citep[e.g.,][]{SCM16,ZDX18,LTR18}.

\subsubsection{Disc-dominated galaxies} \label{sec:Results:Size:LTGs}
\begin{figure*}
\centering \includegraphics[width=\figdblsize]{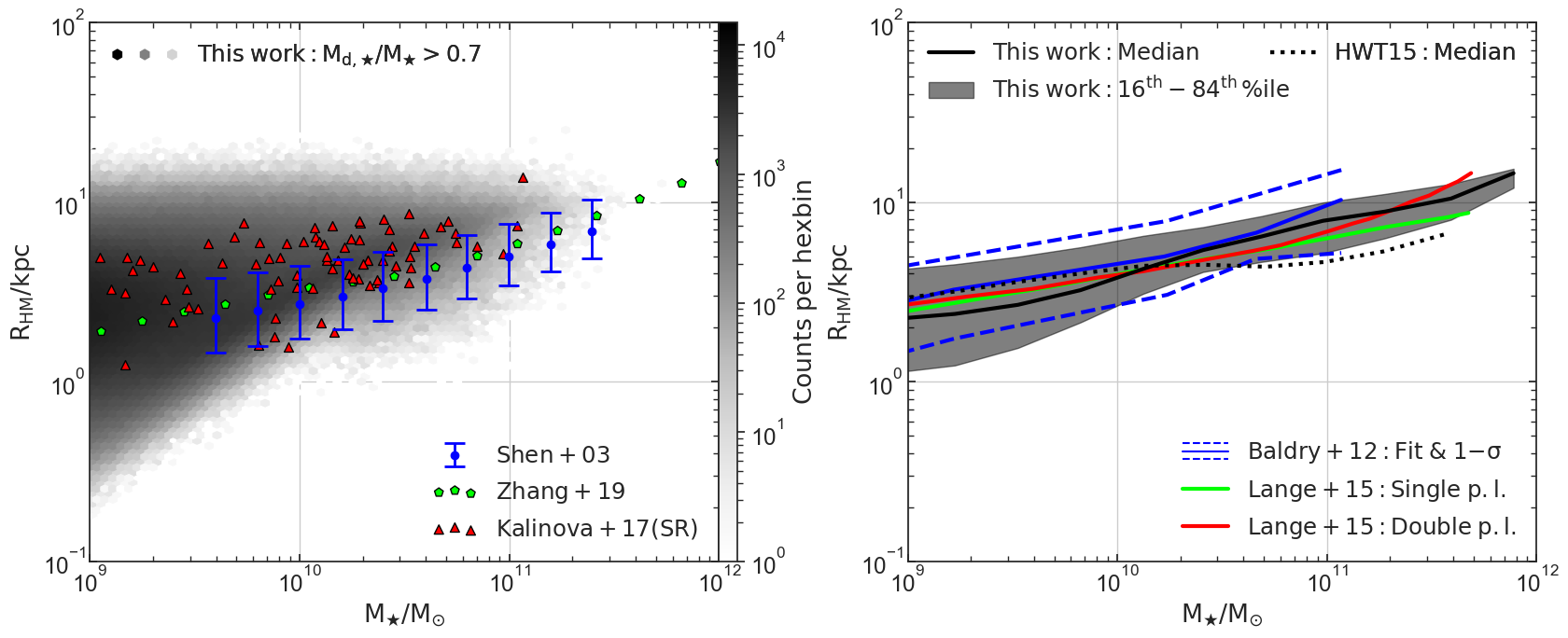}
\caption{Mass-size relation for disc-dominated galaxies at redshift $z \sim
  0.1$. \textbf{Left panel:} Total stellar mass versus stellar half-mass radius
  compared with \protect\cite{SMW03,ZY19} and
  \protect\cite{KCR17}. \textbf{Right panel:} The median and
  $\mathrm{16^{th}-84^{th}}$ percentile range (black shaded region) of the aforementioned relation
  compared with \protect\cite{BDL12} and \protect\cite{LDR15} fit
  lines. Black solid and dotted lines show results from this work and HWT15, respectively.}
\label{fig:SM_Vs_SHMR_LTGs}
\end{figure*}

In \Fig{SM_Vs_SHMR_LTGs} we present the stellar half-mass radius as a function of the total stellar mass for disc-dominated galaxies. In this work we define them as those that have $\Mstar / M_{\star} > 0.7$. We compare our galaxies with the following works:
\begin{itemize}
\item \cite{SMW03}: selected galaxies with concentration index ($c \equiv R_{90}/R_{50})$ $c< 2.86$ from 140,000 SDSS\,DR1
  \citep{YAA00} galaxies at $z < 0.3$.
\item \cite{ZY19}: selected 424,363 galaxies with $c <2.85$ from the New York
  University Value-Added Galaxy Catalog at $z < 0.2$ \citep{BSS05}.
\item \cite{KCR17}: selected slow-rising class galaxies (akin to late-type)
  based on the shapes and amplitude of the circular velocity curve of 238 CALIFA
  galaxies at $z < 0.03$ \citep{FLV17}.
\item \cite{BDL12}: selected late-type galaxies based on colour-magnitude
  diagrams of 5,210 GAMA galaxies at $z<0.06$ \citep{DHK11}.
\item \cite{LDR15}: selected late-type galaxies by visually classifying GAMA\,II
  galaxies in the redshift range $0.01<z<0.1$ \citep{LBD15}.
\end{itemize}

As explained in \Sec{Model:Discs}, we assume that the cold gas loses a fraction
of its specific angular momentum to the dark matter halo during the cooling
process, hence the cold gas discs are expected to be more compact than those
produced by HWT15: this trait is then inherited by the stellar discs. This
behaviour is present at the low mass end of \Fig{SM_Vs_SHMR_LTGs}, however for
stellar masses above $\sim 10^{10} \Msun$ disc instabilities begin to
redistribute stellar material between the disc and the bulge and create a
significant population of pseudo-bulges, as indicated in panel (2,1) of
\Fig{SM_Decomp}. This mechanism causes the expansion of the
disc \footnote{During instabilities low angular momentum material is moved
  inwards and higher angular momentum material is transferred outwards. Hence, while the inner parts of the disc grow denser, the outer parts expand and become more diffuse (see \Sec{Model:DI}).} and produces discs larger than the HWT15 at intermediate and higher masses (see also \Sec{A:AM}). For those reasons, our results show a steeper relation that is in better agreement with the observational data and provide a significant improvement over past modelling attempts \citep[e.g., top panel of Figure 2 of][]{GWB11}.

\subsubsection{Bulge-dominated galaxies} \label{sec:Results:Size:ETGs}
\begin{figure*}
\centering
\includegraphics[width=\figdblsize]{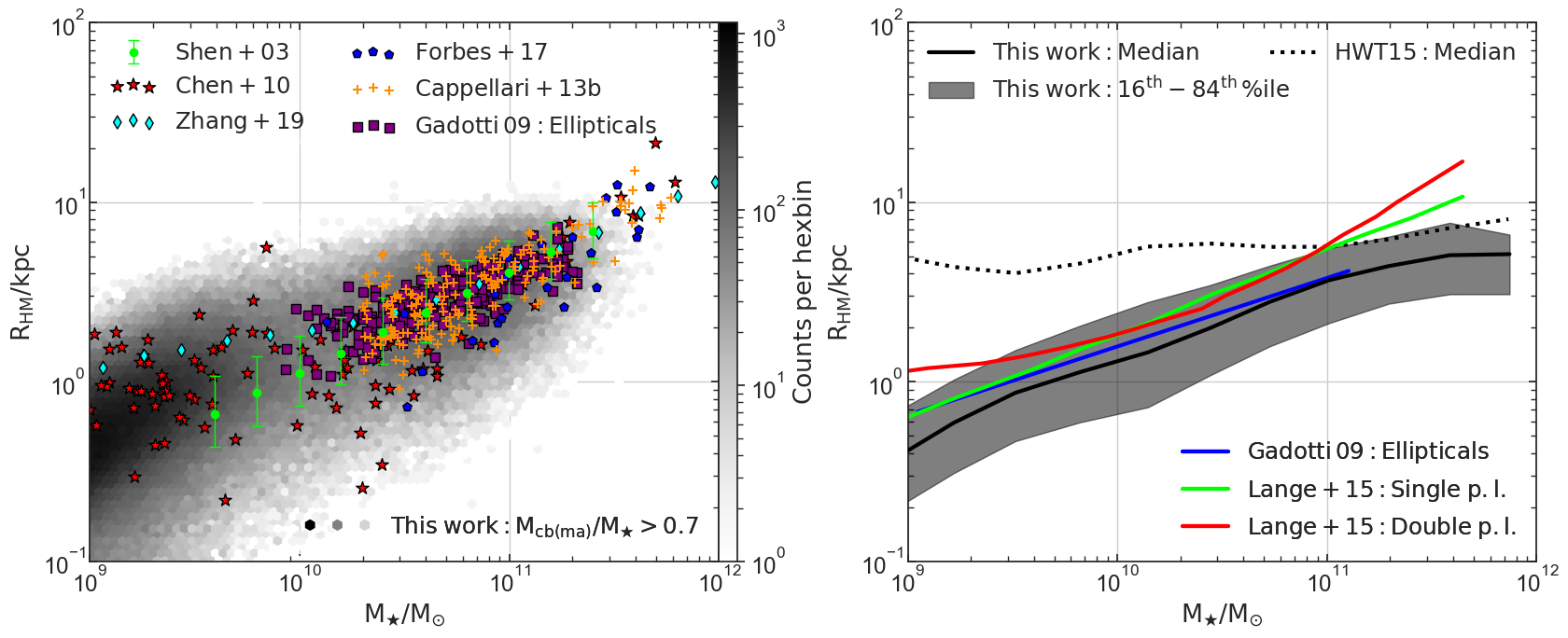}
\caption{Mass-size relation for bulge-dominated galaxies at redshift $z \sim
  0.1$. \textbf{Left panel:} Total stellar mass versus stellar half-mass radius compared with
  \protect\cite{SMW03,CCW10,ZY19,FSS17,CMA13} and
  \protect\cite{G09}. \textbf{Right panel:} The median and
  $\mathrm{16^{th}-84^{th}}$ percentile range (black shaded region) of the aforementioned relation
  compared with \protect\cite{G09} and
  \protect\cite{LDR15} fit
  lines. Black solid and dotted lines show results from this
  work and HWT15, respectively.}
\label{fig:SM_Vs_SHMR_ETGs}
\end{figure*}

The HWT15 model gives sizes of bulge-dominated galaxies that are too large for a given
mass. That motivated us to introduce energy dissipation in gas rich mergers, as described in \Sec{Model:Classical}. The result of that is shown in
\Fig{SM_Vs_SHMR_ETGs} where we show the total stellar mass versus stellar
half-mass radius for galaxies with $M_\mathrm{cb(ma)} / M_{\star} > 0.7$. This sample contains galaxies which composed most of their stellar mass through major mergers, akin to ellipticals. We compare with the following observational datasets:
\begin{itemize}
\item \cite{SMW03}: selected galaxies with $c > 2.86$ from 140,000 SDSS\,DR1
  \citep{YAA00} galaxies at $z<0.3$.
\item \cite{CCW10}: selected about 100 early-type galaxies that populate the red
  sequence in the Virgo cluster from SDSS\,DR5 \citep{AAA07}.
\item \cite{ZY19}: selected 424,363 galaxies with $c > 2.85$ from the New York
  University Value-Added Galaxy Catalog at $z < 0.2$ \citep{BSS05}.
\item \cite{FSS17}: selected galaxies from the SLUGGS survey which targeted 25
  nearby $(\mathrm{D \leq 25\,Mpc})$ massive early-type galaxies in different
  environments \citep{BRS14}.
\item \cite{CMA13}: selected 260 early-type galaxies from the
 $\mathrm{ATLAS^{3D}}$ project at $z = 0$ \citep{CEK11}.
 \item \cite{G09}: selected galaxies with $c > 2.5$ from the SDSS\,DR2 \citep{AAA04}.
\item \cite{LDR15}: selected early-type galaxies by visually classifying
  GAMA\,II galaxies in the redshift range $0.01<z<0.1$ \citep{LBD15}.
\end{itemize}

The updated merger remnant size recipe introduced in this work gives more
compact remnant sizes at the low-mass end compared to HWT15 which
over-predicted the size of the smallest galaxies. We can clearly see that our
median line agrees well with a single power law for masses below $10^{10}\Msun$,
as indicated by \cite{LDR15}. However, at the high mass end we do not see the sharp upturn in size indicated by their double power-law model. We note that there is an increase in intracluster light in the most massive haloes that we do not include when calculating the size of the central galaxies.

\subsubsection{The dependence of disc scale length on morphology}
\label{sec:Results:Size:DSL}
\begin{figure*}
\centering \includegraphics[width=\figdblsize]{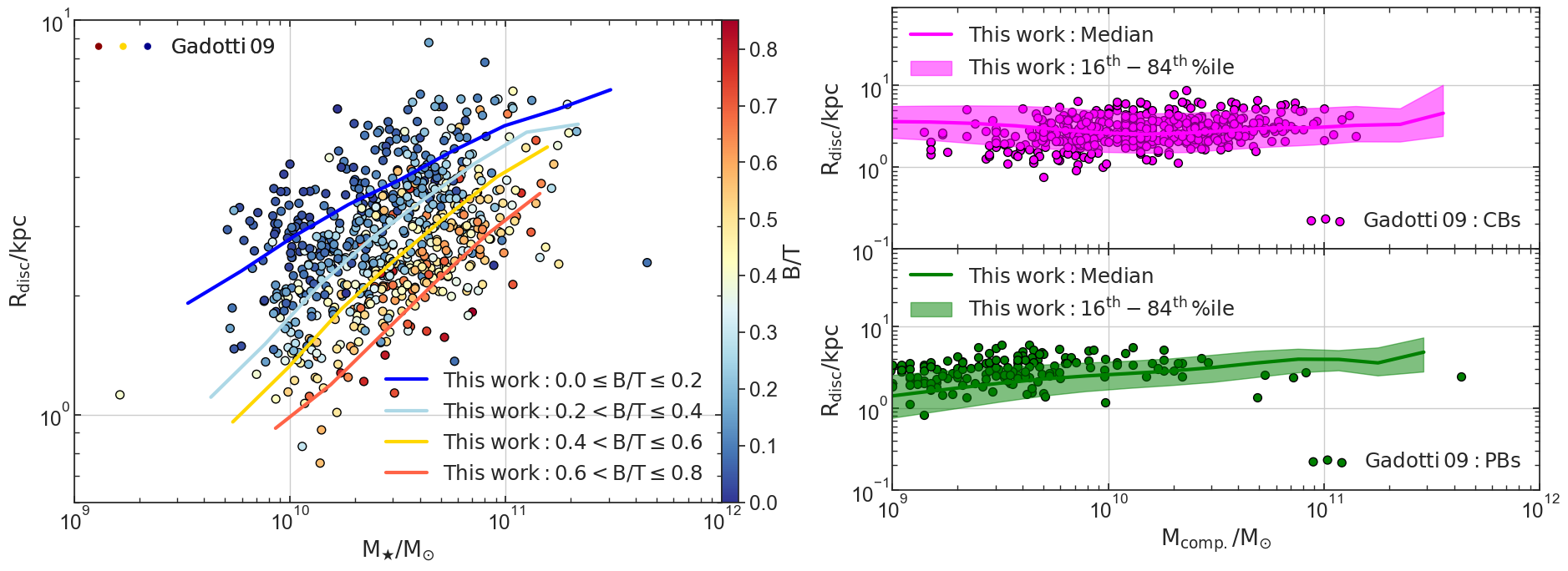}
\caption{Disc scale length as a function of mass at redshift $z \sim 0.05$. \textbf{Left
    panel:} Total stellar mass versus disc scale length compared with
  \protect\cite{G09} galaxies colored by different $B/T$ ratios. \textbf{Right
    panel:} Classical and pseudo-bulge mass against disc scale length compared
  with \protect\cite{G09} classical and pseudo-bulges, respectively.}
\label{fig:M_Vs_DSL}
\end{figure*}

In \Fig{M_Vs_DSL} we present three different versions of the disc scale length versus mass relation. The left panel contains our median lines for 4 different bulge-to-total stellar mass ratios and \cite{G09} galaxies color-coded by their $B/T$ luminosity ratio. The \lgal\ model shows adequate agreement with the observed behaviour at all masses, which indicates that the disc scale lengths
decrease as the $B/T$ ratio increases.

For the right panel we selected galaxies with classical bulges through minor mergers and galaxies with pseudo-bulges and plotted their disc scale lengths as a function of the mass of these bulges. \cite{G09} fitted different profiles in each galaxy image in his sample and used a bulge profile which is described by a \cite{S68} function; when n = 4 the profile is a \cite{V48}, while n = 1 corresponds to an exponential bulge (i.e.,pseudo-bulge). We find a strong agreement with the observed trends which suggests that, as expected, galaxies with more extended discs tend to host more massive pseudo-bulges. This slope appears to be steeper for galaxies with pseudo- instead of classical bulges and as we discussed in \Sec{Model:DI} bar formation is expected to expand the outer parts of the disc \citep[e.g.,][]{KK04}.

\section{Conclusions} \label{sec:Conc}
This paper addresses some deficiencies in the otherwise very successful
\cite{HWT15} \SAM\ with regard to bulge formation via disc instabilities and merger remnant sizes. In making the latter change, we drew inspiration from the work of \cite{CDC07,CPP11} and \cite{TMC16}. The main changes are:
\begin{itemize}
\item the specific angular momentum of accreted gas is reduced to 0.8 times that
  of the dark matter halo;
\item an improved disc instability recipe that considers both the gas and the
  stars, rather than just the latter;
\item the introduction of dissipation in gas-rich mergers.
\end{itemize}

\noindent Our main conclusions are as follows:
\begin{itemize}
\item The updated disc instability recipe allows us to have an impressive
  agreement with the observed fraction of different galactic morphologies and the stellar mass functions of different galactic types. 
\item The stellar half-mass radius and specific angular momentum of disc-dominated galaxies is in
  great agreement with the observed relations due to the reduction of the
  initial angular momentum of the gas disc.
\item Highly dissipative mergers result in more compact remnants which match the observed mass-size relation of bulge-dominated galaxies.
\item The tight relation between the stellar disc scale length and mass is still
  present after the assumption that the gas loses 20$\%$ of its initial
  specific angular momentum during cooling.
\end{itemize}

\section*{Acknowledgements}
The authors contributed in the following way to this paper. DI undertook the vast majority of the coding and data analysis, and produced the figures. PAT supervised DI and together they drafted the paper. BMH, MTS and JMH provided input on content and helped with interpretation of the results. All authors helped to proof-read the text.

The authors would like to thank the anonymous referees for their useful comments which contributed to improve this work. DI (ORCID 0000-0003-2946-8080) would like to acknowledge his family members for their financial support. PAT (ORCID 0000-0001-6888-6483) acknowledges support from the Science and Technology Facilities Council (grant number ST/P000525/1). The work of BMH (ORCID 0000-0002-1392-489X) was supported by Advanced Grant 246797 ``GALFORMOD” from the European Research Council and by a Zwicky Fellowship. MTS acknowledges support from a Royal Society Leverhulme Trust Senior Research Fellowship (LT150041)

\bibliographystyle{mnras}
\bibliography{mn-jour,di}

\appendix
\section{Sanity checks} \label{A}
\subsection{Angular momentum} \label{sec:A:AM}
\begin{figure}
\centering \includegraphics[width=\figsize]{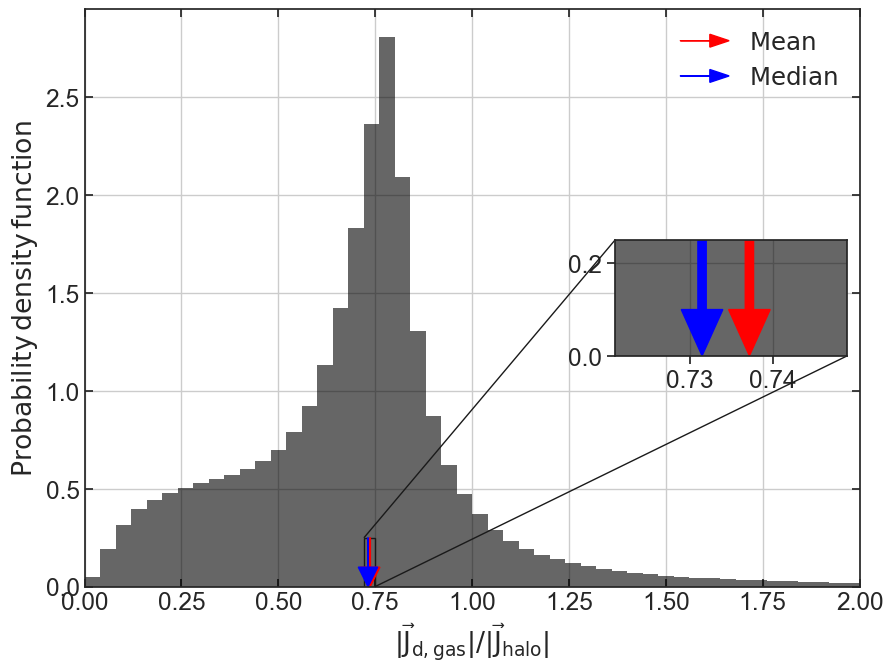}
\caption{Probability density function of the ratio between the specific angular momentum of the gas disc and that of the halo at redshift z $\sim$ 0.0. The red (blue) arrow indicates that the mean (median) value for the galaxies in our sample is 0.737 (0.731).}
\label{fig:PDF}
\end{figure}

As discussed in \Sec{Intro:PrevModel:AM} the assumptions under which the \lgal\ and the majority of SAMs calculate disc sizes and specific angular momenta have been criticised by simulators who studied the connection between the dark matter halo and its baryons. These studies have found that the specific angular momentum of the latter is notably lower than that of the former \citep[e.g.,][]{KG91,NW93,NW94,NFW95,NS97,CLB00,KMW07,ZOF07,KDS11,DDH15,SLC17}. This motivated us to include the factor $f$ in \Eq{DeltaJgas} and, as discussed in \Sec{Model:Discs}, in this work we assumed that during cooling the gas disc loses 20 per cent of its specific angular momentum.

In \Fig{PDF} we show the probability density function of the ratio between the specific angular momentum of the gas disc and that of the halo. In the L-Galaxies model we follow the changes in the total angular momentum vector of the gas disc during each time-step from a variety of physical processes, as \Eq{DeltaJgas} denotes. Hence, the current value of that ratio represents the angular momentum accumulated over cosmic history and does not directly measure the instantaneous rate of accretion of angular momentum. Thus, even though $f$ equals 0.8 for cooling gas, a slight bias to lower specific angular momenta and a galaxy-to-galaxy scatter emerges from the other mechanisms that affect the angular momentum of each galaxy.

\begin{figure}
\centering \includegraphics[width=\figsize]{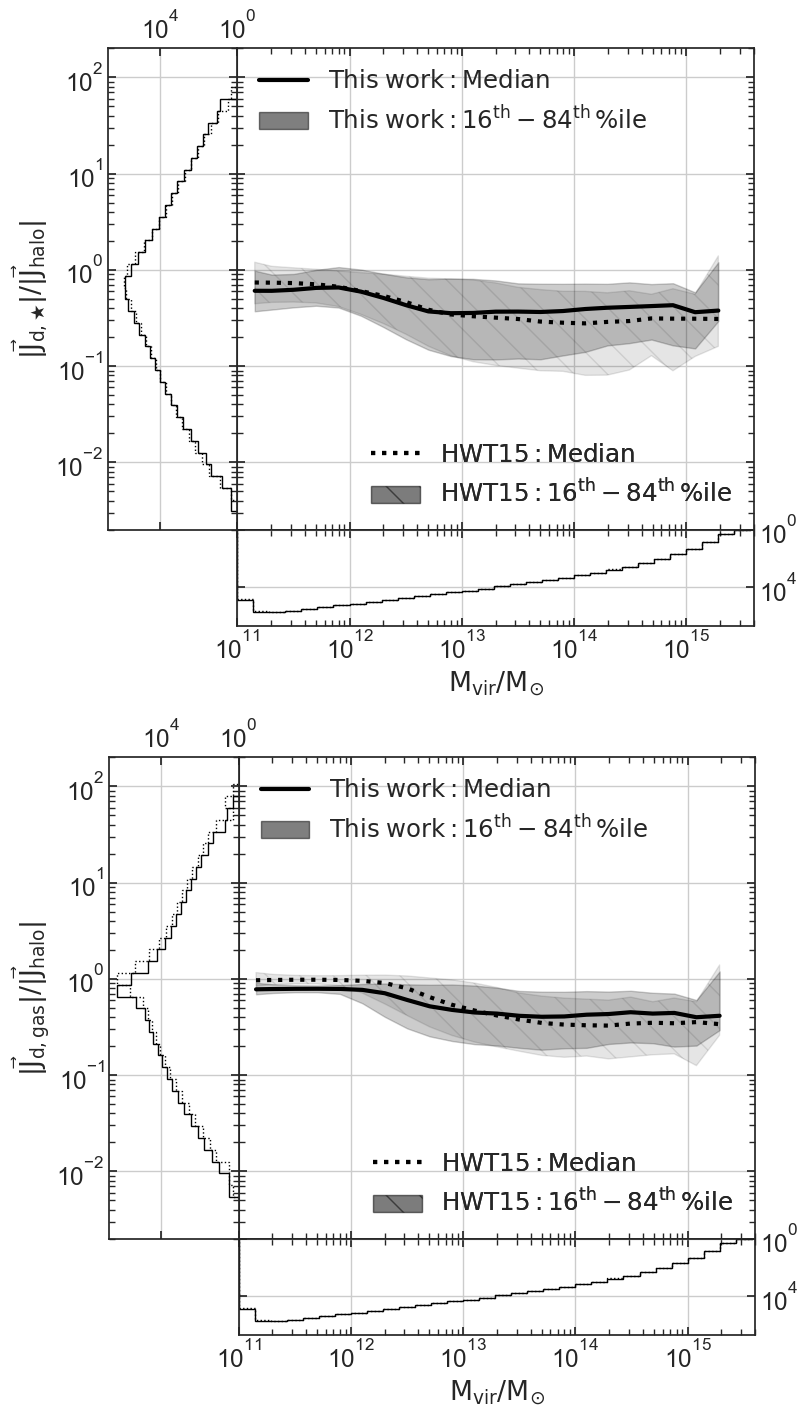}
\caption{Baryonic-to-halo specific angular momentum ratio as a function of halo mass at redshift z $\sim$ 0.0 \textbf{Top panel:} The median and
  $\mathrm{16^{th}-84^{th}}$ percentile range of the stellar-to-halo specific angular momentum as a function of halo mass.
  \textbf{Bottom panel:} The median and
  $\mathrm{16^{th}-84^{th}}$ percentile range of the gas-to-halo specific angular momentum as a function of halo mass. In both panels the panels attached to the axes are the histograms of the corresponding property. Black solid lines and shaded regions show results from this work, while black dotted lines and diagonally hatched regions from HWT15.}
\label{fig:Mvir_Vs_AMRatio}
\end{figure}

Furthermore, we investigate how the baryonic specific angular momentum relates to halo properties. \Fig{Mvir_Vs_AMRatio} shows the ratio of stellar-to-halo and gas-to-halo specific angular momentum as a function of halo mass. Our results indicate that the stellar specific angular momentum is lower than that of the cold gas, which is in turn slightly lower than that of the halo. This behaviour is consistent with the one found by previous simulations \citep{TRD15,JDK18}. We find a slight decrease in the aforementioned ratios as halo mass increases (see also \Sec{A:SD} for the effect of disc instabilities and angular momentum losses on discs), which is in broad agreement with the trends reported in recent theoretical studies \citep[e.g.,][]{PPF18}. We also notice that the scatter seen in \Fig{PDF} is also prominent in the y-axis histogram in the bottom panel of \Fig{Mvir_Vs_AMRatio} where the use of a log scale for the normalisation shows that it spans a wide range of values.

\subsection{Stellar disc} \label{sec:A:SD}
\begin{figure}
\centering \includegraphics[width=\figsize]{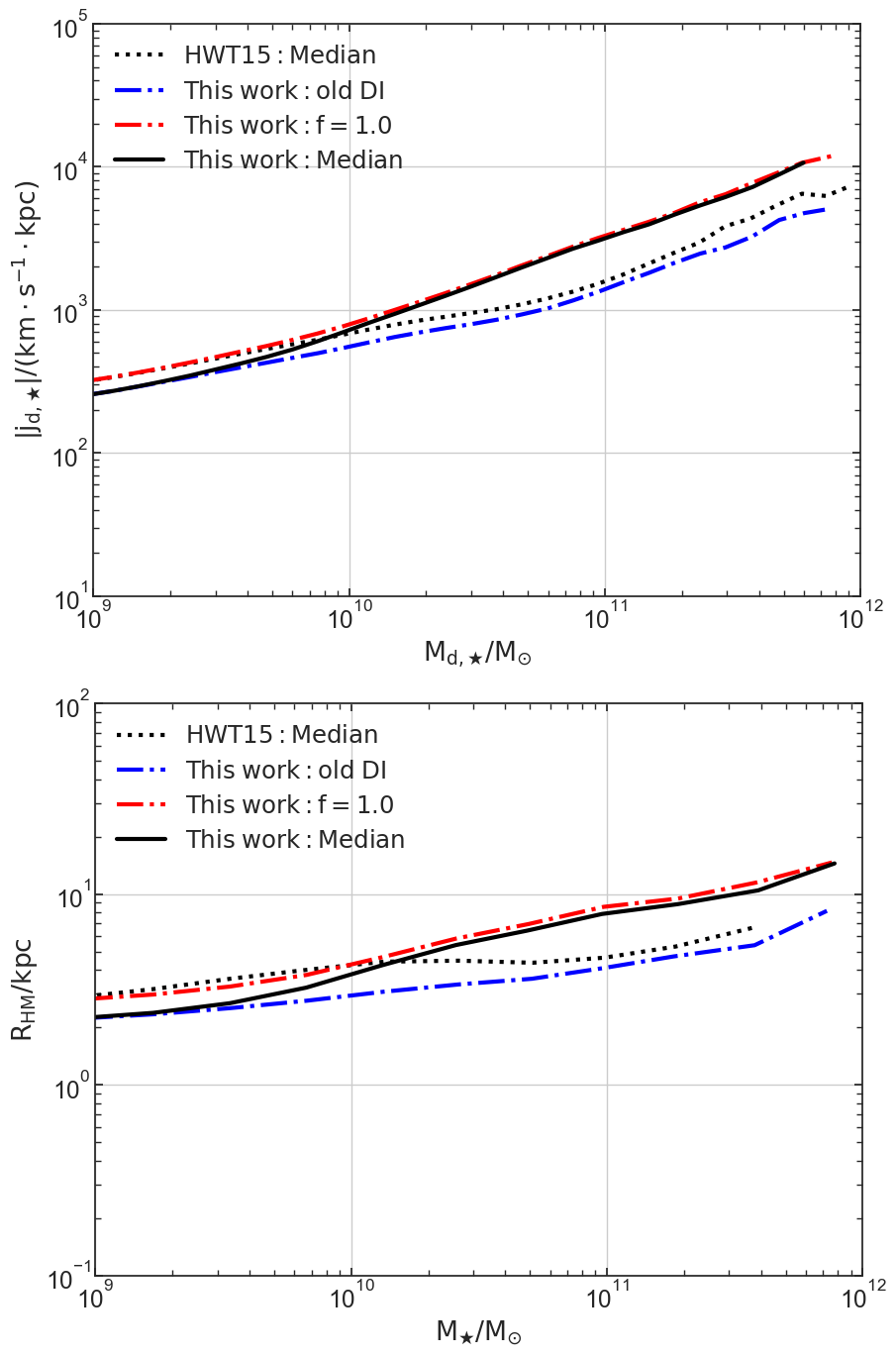}
\caption{ \textbf{Top panel:} Stellar mass versus specific angular momentum for disc-dominated galaxies at redshift z $\sim$ 0.0. \textbf{Bottom panel:} Mass-size relation for disc-dominated galaxies at redshift $z \sim 0.1$. In both panels solid and dotted black lines show results from this work and HWT15, respectively, and the dashdotted red and blue lines show our model (i.e., new disc instability recipe) with f=1.0 and our model (i.e., f=0.8) with the old disc instability recipe (i.e., from HWT15), respectively.}
\label{fig:SD}
\end{figure}

In this section we reproduce \Fig{DM_Vs_DS} and \Fig{SM_Vs_SHMR_LTGs} for different flavours of the \lgal\ model in order to evaluate the effect of disc instabilities and angular momentum losses on stellar discs. In \Fig{SD} we show the median lines for the mass versus specific angular momentum (top panel) and the mass-size relation (bottom panel) for this version of the model, the HWT15, this model (i.e., with the disc instability recipe described in \Sec{Model:DI}) with f=1.0 and this model (i.e., with the angular momentum losses described in \Sec{Model:Discs}) with the old disc instability recipe (i.e., from HWT15). In both panels we see that at the low mass end ($<10^{10}\Msun$) the solid black and the blue line converge since angular momentum losses produce more compact and slowly rotating discs. On the other hand, for stellar masses $\sim 10^{10}\Msun$ and above, where the formation of pseudo-bulges is expected to happen (see panel (2,1) in \Fig{SM_Decomp}), we see a drastic change in the properties of stellar discs since disc instabilities increase their size and specific angular momentum. Hence, the solid black line now follows closely the red one. In general, the conclusions drawn from \Fig{SD} support the arguments put forward in \Sec{Results:Spin} and \Sec{Results:Size:LTGs}, where we argued that the steepening of the median line at the high mass end in \Fig{DM_Vs_DS} and \Fig{SM_Vs_SHMR_LTGs} is due to our new instability recipe. 

\subsection{Stellar mass function}
\begin{figure}
\centering \includegraphics[width=\figsize]{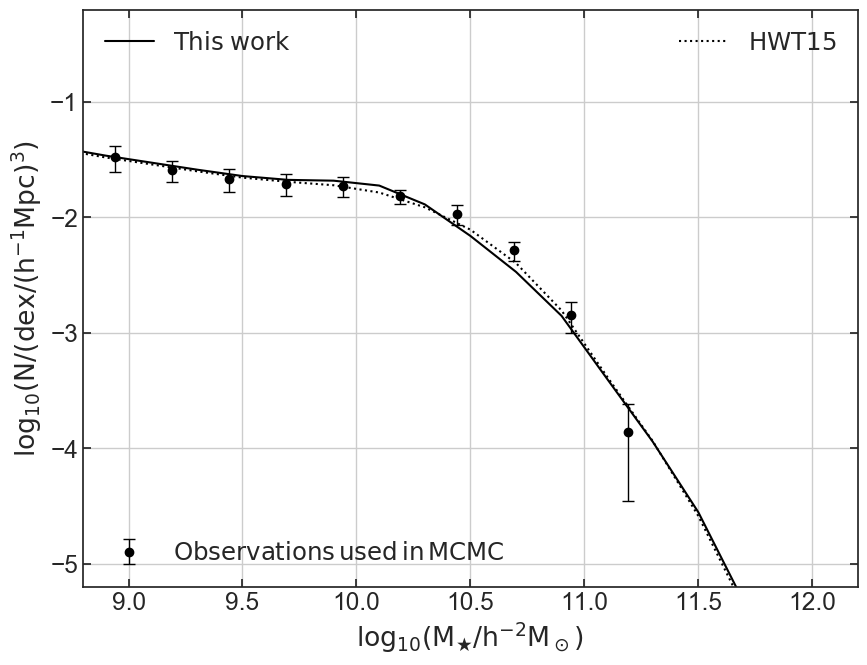}
\caption{Total stellar mass functions at redshift z $\sim$ 0.0. Solid and dotted lines show results from this work and HWT15, respectively, and black circles represent the combined observational data used to constrain the MCMC in HWT15.}
\label{fig:SMF}
\end{figure}

\Fig{SMF} shows the total stellar mass function. The black circles represent the observational data used by HWT15 to constrain the MCMC. Instead of running a new MCMC analysis and readjusting the free parameters of the \lgal\ model, we chose to follow HWT15 results. Hence, even though we have significantly altered the \lgal\ model we see that our stellar mass function is in close agreement with the one produced by them. 

\subsection{Black hole-bulge mass relation}
\label{sec:Results:BH}

\begin{figure}
\centering
\includegraphics[width=\figsize]{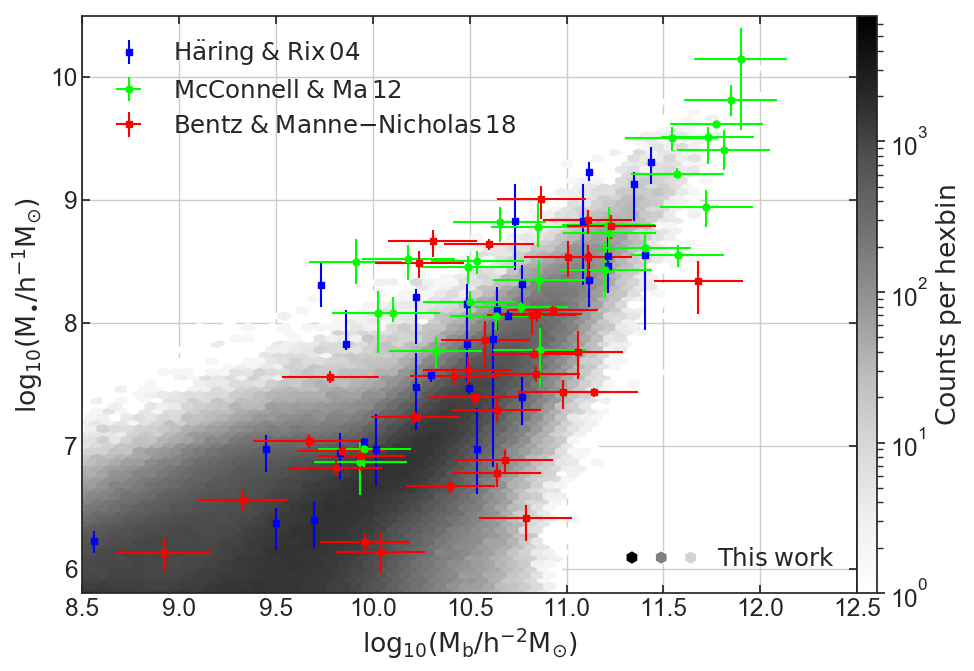}
\caption{Black hole-bulge mass relation at redshift z $\sim$ 0.0. Black hexagons
  represent our galaxies; blue, green and red circles are observations from
  \protect\cite{HR04}, \protect\cite{MM13} and \protect\cite{BM18},
  respectively.}
\label{fig:BM_Vs_BHM}
\end{figure}

In this work we updated the processes responsible for the growth of bulges via mergers and disc instabilities; where the latter mechanism feeds a percentage of the unstable cold gas into the central super-massive black hole. Hence, \Fig{BM_Vs_BHM} provides a sanity check for our new model since it illustrates that the \lgal\ model is still able to reproduce the tight black hole-bulge mass relation and shows an impressive agreement with the observational data at all masses.

We compare our simulated data with a sample of 30 nearby galaxies introduced by \cite{HR04}, 72 galaxies compiled by \cite{MM13} and 37 galaxies selected by \cite{BM18} from the Hubble Space Telescope images and deep, ground-based near-infrared images. Even though for $10^{9} \Msun < \Mbulge < 10^{10.5} \Msun$ the \lgal\ model predicts a large scatter in black hole masses, the majority of our galaxies form at all masses an almost linear relation in log-space (i.e.~a power-law in linear-space) between black hole and bulge mass, as expected \citep[e.g.~][]{BCC12,G12,MM13}.

\section{Supplementary results} \label{B}
\subsection{The Tully-Fisher relation}
\label{sec:Results:Tully-Fisher}
\begin{figure}
\centering \includegraphics[width=\figsize]{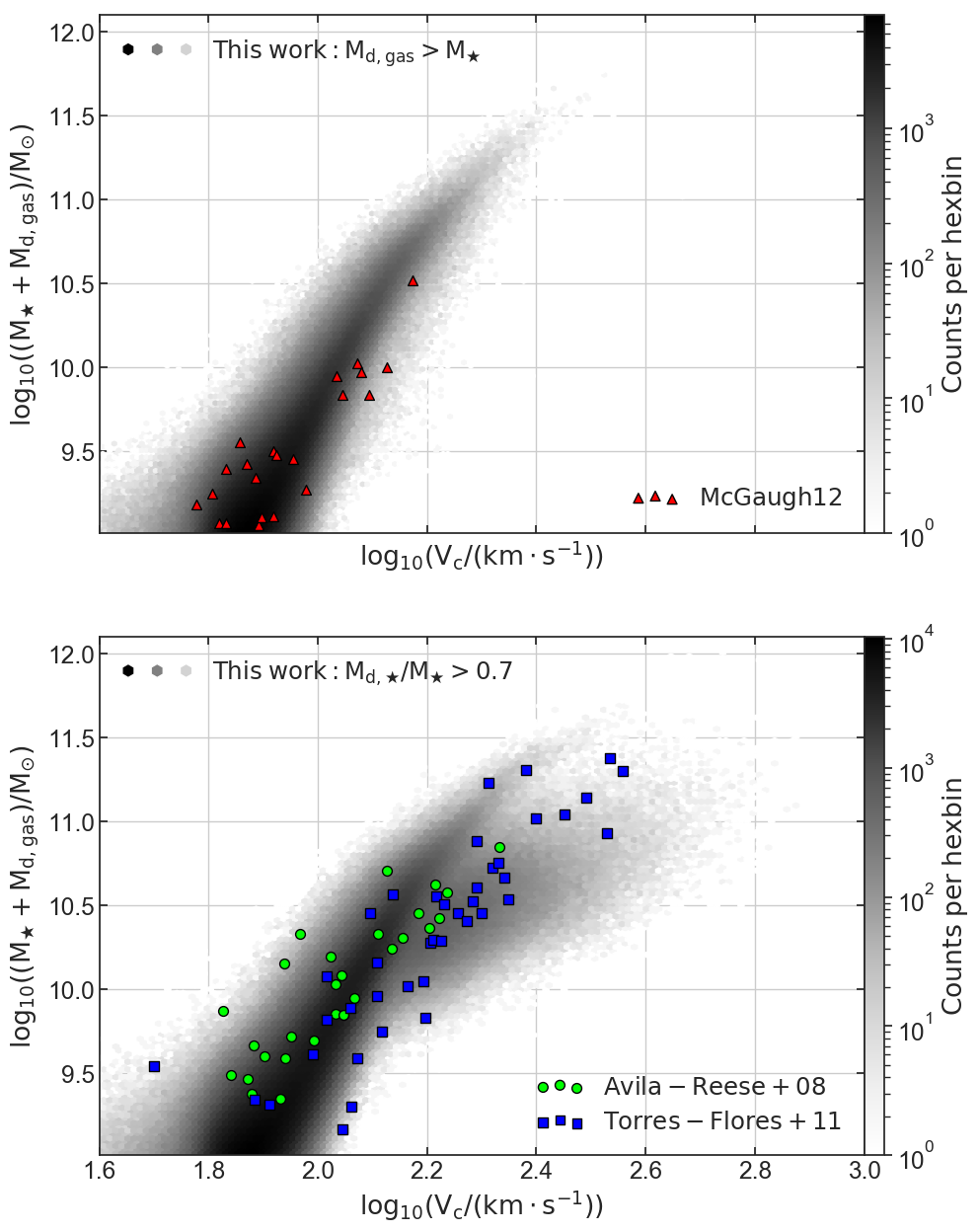}
\caption{Baryonic Tully-Fisher relation at redshift $z \sim 0.0$. \textbf{Top panel:}
  Gas-dominated galaxies compared with a dataset from
  \protect\cite{M12}. \textbf{Bottom panel:} Disc-dominated galaxies compared
  with \protect\cite{AZF08} and \protect\cite{TEA11}.}
\label{fig:TF}
\end{figure}

The Tully-Fisher relation \citep{TF77} describes an empirical correlation
between the intrinsic luminosity and the emission line width of rotating spiral galaxies. A more useful form for our purposes has been proposed by \cite{MSB00} that relates the total baryonic mass and the rotation velocity.

In this work we adopt, for simplicity, as the typical rotation velocity for both
the gaseous and the stellar disc, the maximum circular velocity of the
surrounding dark matter halo ($V_\mathrm{max}$). This assumption is in agreement with \cite{TWP10} who found that the maximum circular velocities of dark matter haloes are very similar to the maximum rotation velocities of discs. In the top panel of \Fig{TF} we compare gas-dominated (i.e., $\Mgas > M_{\star}$) galaxies produced by the \lgal\ model with the dataset used by \cite{M12} which consist of gas dominated galaxies from \cite{BCK08,SMS09} and \cite{TBM09}. Furthermore, in the bottom panel we investigate the baryonic Tully-Fisher relation of disc-dominated galaxies (i.e., $\Mstar / M_{\star} > 0.7$) where we include observations from \cite{AZF08} \citep[normal, non–interacting disc galaxies compiled from the literature and homogenized in][]{ZAH03} and \cite{TEA11} \citep[spiral and irregular galaxies from Gassendi HAlpha survey of SPirals, GHASP][]{EAM08a,EAM08b}.

As shown in both panels of \Fig{TF}, our galaxies follow a tight relation that
is in close agreement with the observational data. However, in the bottom panel we notice that some of the galaxies with the highest circular velocities appear to be less massive than those observed by \cite{AZF08} and \cite{TEA11}. This results from the fact that for baryonic masses $\mathrm{log}_{10}((M_\mathrm{\star} + M_\mathrm{d,gas})/\Msun) > 10$ our galaxies split into two groups, where the lower one represents extremely gas poor quiescent galaxies whose contribution to the total baryonic mass is not significant.

\subsection{A mass-mass relation} 
\label{sec:Results:MM}
\begin{figure}
\centering \includegraphics[width=\figsize]{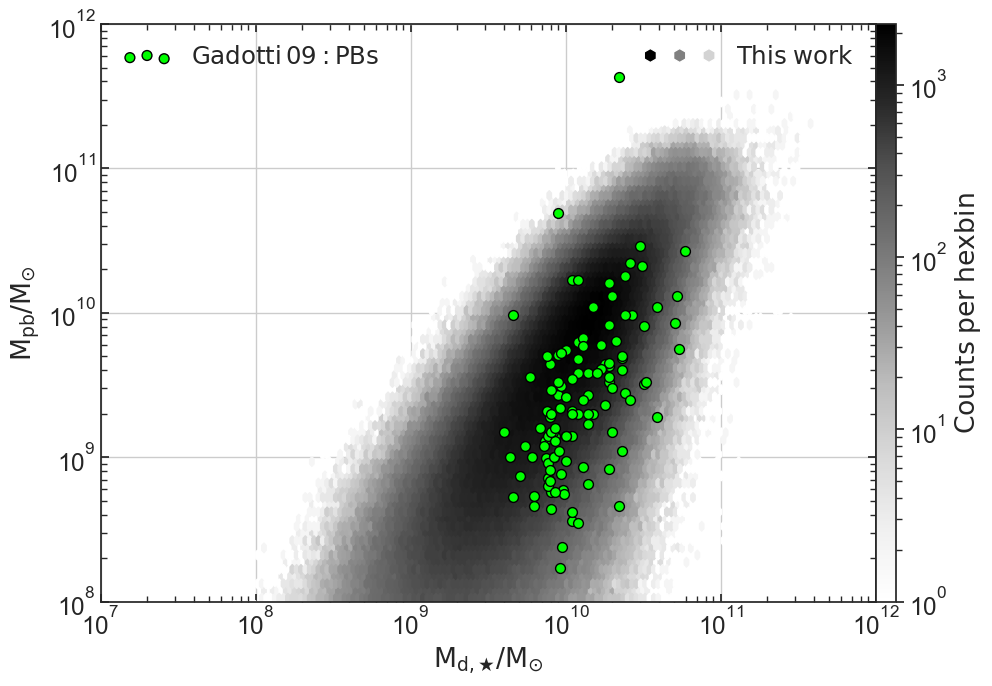}
\caption{Pseudo-bulge mass as a function of disc mass at redshift $z \sim 0.0$ compared with \protect\cite{G09} data.}
\label{fig:MM}
\end{figure}

In the \cite{ELN82} criterion, more massive discs are more unstable. Hence, we expect a tight relation between the disc mass and pseudo-bulge mass to be present in our model. Interestingly, a similar trend appears to exist in real galaxies.

In \Fig{MM} we plot the pseudo-bulge mass as a function of the disc mass. Our data suggest that since more massive discs are more unstable they will be able to create more massive pseudo-bulges. In the \cite{G09} galaxies the same trend is observed as more massive discs host more massive pseudo-bulges. However, their slope appears to be slightly steeper than the one produced by the \lgal\ model.

\label{lastpage}
\end{document}